\documentstyle[epsfig,twocolumn]{mn}

\title[An updated GRBs Hubble diagram]{An updated Gamma\-\,Ray Bursts Hubble diagram}
\author[V.F. Cardone et al.]{V.F. Cardone$^{1,2}$, S. Capozziello$^{1,3}$, M.G. Dainotti$^{4,5,6}$\\
$^1$ Dipartimento di Scienze Fisiche, Universit\`{a} di Napoli "Federico II",
Complesso Universitario di Monte Sant' Angelo, \\ Edificio N, via Cinthia,
80126 - Napoli, Italy \\ $^2$ Dipartimento di Fisica Generale, Universit\`{a} di Torino,
and I.N.F.N. - Sez. di Torino, Via Pietro Giuria 1, 10125 - Torino, Italy \\
$^3$ I.N.F.N., Sez. di Napoli, Complesso
Universitario di Monte Sant' Angelo, Edificio G, via Cinthia, 80126 -
Napoli, Italy \\ $^4$ICRANet and ICRA, Piazzale della Repubblica 10, 65122 - 
Pescara, Italy \\ $^5$ Dipartimento di Fisica, Universit\`{a} di Roma "La
Sapienza", Piazzale Aldo Moro 5, 00185 - Roma, Italy \\
$^6$ Observatorium Astronomiczne, Universytet Jagiellonski, ul. Orla 175 - 31-501 Krakow, Poland}

\date{Accepted xxx, Received yyy, in original form zzz}

\begin{document}
\maketitle

\begin{abstract}
Gamma ray bursts (GRBs) have recently attracted much attention as a
possible way to extend the Hubble diagram to very high redshift. To this
aim, the luminosity (or isotropic emitted energy) of a GRB at redshift $z$
must be evaluated from a correlation with a distance independent quantity
so that one can then solve for the luminosity distance $D_L(z)$ and hence
the distance modulus $\mu(z)$. Averaging over five different two parameters
correlations and using a fiducial cosmological model to calibrate them,
Schaefer (2007) has compiled a sample of 69 GRBs with measured $\mu(z)$
which has since then been widely used to constrain cosmological parameters.
We update here that sample by many aspects. First, we add a recently found
correlation for the X\,-\,ray afterglow and use a Bayesian inspired fitting
method to calibrate the different GRBs correlations known insofar assuming
a fiducial $\Lambda$CDM model in agreement with the recent WMAP5 data.
Averaging over six correlations, we end with a new GRBs Hubble diagram
comprising 83 objects. We also extensively explore the impact of varying
the fiducial cosmological model considering how the estimated $\mu(z)$
change as a function of the $(\Omega_M, w_0, w_a)$ parameters of the
Chevallier\,-\,Polarski\,-\,Linder phenomenological dark energy equation of
state. In order to avoid the need of assuming an {\it a priori}
cosmological model, we present a new calibration procedure based on a model
independent local regression estimate of $\mu(z)$ using the Union SNeIa
sample to calibrate the GRBs correlations. This finally gives us a GRBs
Hubble diagram made out of 69 GRBs whose estimated distance modulus
$\mu(z)$ is almost independent on the underlying cosmological model.

\end{abstract}


\section{Introduction}

The Hubble diagram of Type Ia Supernovae (SNeIa) provided the first
evidence of the cosmic speed up \cite{SCP97,HZT98} opening the way to a
flood of theoretical and observational efforts devoted to investigate the
nature and the nurture of such a phenomenon. The astrophysical data coming
from later SNeIa samples \cite{Riess07,SNLS,ESSENCE,,Davis07,Union},
Baryonic Acoustic Oscillations \cite{E05,P07}, the anisotropy spectrum of
cosmic microwave background radiation \cite{Boom,WMAP,WMAP,WMAP5} and the
large scale structure data from large galaxy redshift surveys
\cite{D02,P02,Sz03,H03} have strenghtened the then surprising result. As a
consequence, there is nowadays little doubt that the universe is spatially
flat, has a low matter content ($\Omega_M \simeq 0.3$) and is undergoing a
phase of accelerated expansion. To close the energy budget and drive the
cosmic speed up, a new fluid with negative pressure dubbed dark energy has
been invoked as the only viable solution. Understanding what dark energy
really is stands out as one of the most fascinating problems of modern
cosmology. Roughly speaking, we can divide the different proposals in two
broad categories. In the first class, one can include all models
considering dark energy as a true fluid with unusual properties with
proposals ranging from the classical cosmological constant $\Lambda$
\cite{Carroll,Sahni} to scalar fields running down their self interaction
potential \cite{PR03,Pad03,CST06}. On the other hand, one may give off
General Relativity resorting to modified gravity theories such as $f(R)$
models (see, e.g., Capozziello et al. 2008, Sotiriou \& Faraoni 2008 and
references therein) and braneworld cosmologies \cite{DGP00,LSS04}
manifesting themselves as a fictitious dark energy\,-\,like fluid.

Notwithstanding their radically different inspiring phylosophies, all the
viable theories share the same ability to be in satisfactory agreement with
the available data. Such a confusing situation is a clear evidence that
present dataset are unable to discriminate among different models. Such a
problem comes out as a consequence of the still low redshift range probed.
The SNeIa Hubble diagram is an illuminating example. As well known, the
distance modulus $\mu(z)$ scales linearly with the logarithm of the
luminosity distance $d_L(z)$ which depends on the dark energy equation of
state through a double integration. Discriminating among different models
asks for extending the Hubble diagram to higher redshifts since the
expressions for the $H(z)$ and hence $\mu(z)$ are more and more different
as one goes to higher and higher $z$ values.

Unfortunately, such a project can not be undertaken relying on SNeIa since,
even with excellent space based projects such as SNAP \cite{SNAP}, it is
almost impossible going to $z > 2$. Indeed, up to now, most of the SNeIa
observed have redshift $z \le 1.2$ with only one SN at $z = 1.7$. On the
contrary, Gamma Ray Bursts (GRBs) are ideal candidates from this point of
view. Indeed, thanks to their enormous energy release, they have been
observed up to redshift $z > 5$, the most distant one being GRB\,080913
with $z = 6.7$ \cite{Greiner08}. as high as $z \simeq 6.5$. There is,
however, a fierce problem to overcome. Actually, GRBs are not standard
candles so that it is not immediately possible to build their Hubble
diagram. To this aim, one has to look for a correlation between a distance
dependent quantity and a directly observable property. Many of these
correlations have been suggested in the last few years so that the route
towards making GRBs standardizeable candles (in the same way as SNeIa) is
now open.

Actually, various attempts to build a GRBs Hubble diagram have yet been
made. Bradley (2003) first derived the luminosity distances of 9 GRBs with
known redshifts by using two GRBs luminosity relations thus providing the
first GRBs Hubble diagram. Dai et al. (2004) and Xu et al. (2005) have then
proposed a methodology to constrain cosmological parameters using GRBs and
apply it to preliminary samples (made out of 12 and 17 objects
respectively) relying on the Ghirlanda relation. A Bayesian based approach
was instead proposed by Firmani et al. (2005), later improved by the same
authors \cite{Firmani06} advocating the joint use of GRBs and SNeIa.

As a firt summary of these efforts, Schaefer (2007) has compiled a catalog
of 69 GRBs with measured properties entering the five two parameters
correlation then available. Estimating $\mu(z)$ from each single
correlation and performing a weighted average, Schaefer have finally built
the first GRBs Hubble diagram which has later been used in different
cosmological analyses \cite{Wr07,Wang07,Li08,Daly08,CI08}. We update here
the Schaefer work by many aspects. First, we add the recently found
$L_X\,\-\,T_a$ correlation \cite{DCC} with $L_X$ the X\,-\,ray luminosity
at the time $T_a$ and $T_a$ a timescale characterizing the late afterglow
decay. The use of this new correlation allows us to both increase the GRBs
sample and reduce the uncertainty on $\mu(z)$. We then recalibrate all the
six correlations considered using a fiducial $\Lambda$CDM cosmological
model in agreement with the WMAP5 \cite{WMAP5} data and a Bayesian
motivated fitting technique \cite{Dago05}. We are now able to build a new
GRBs Hubble diagram whose dependence on the background cosmological model
used for the calibration is explored. Finally, we present a novel method to
escape the circularity problem, i.e. to calibrate the GRBs empirical laws
without assuming any a priori cosmological model. This gives us then the
possibility to build up a new GRBs Hubble diagram less affected by
systematic uncertainties related to the unknown cosmological scenario.

The plan of the paper is as follows. In Sect. 2, we briefly review the
different GRBs 2D correlations known insofar and present the Bayesian
motivated method we will use in the following to calibrate them. Such a
calibration is performed in Sect. 3.1 using a fiducial $\Lambda$CDM model
with parameters set according to the recent WMAP5 results. These calibrated
relations are then used in Sect. 3.2 to build a Hubble diagram comprising
83 objects, while Sect. 3.3 explores what is the impact of varying the
cosmological model used in the calibration on the estimated distance
modulus. Sect. 4.1 presents the local regression method showing how it is
possible to use the Union SNeIa sample to recover the value of $\mu(z)$ in
a cosmological model independent way. This method is then used to calibrate
the 2D correlations for the GRBs with $z \le 1.4$ in Sect. 4.2 so that a
newly calibrated GRBs Hubble diagram may be built in Sect. 4.3 where we
also compare this latter to the fiducial one. The impact of different kind of 
systematics on the derived Hubble diagrams is investigated in Sect. 5. Having 
thus checked the reliability of the GRBs HDs thus obtained, we present in Sect. 6 
a first cosmological analysis and discusses the importance of GRBs as a cosmological
tool presenting a forecast of the precision affordable on the matter density parameter, 
$\Omega_M$, and the dark energy equation of state. Finally,  Sect. 7 is devoted to the
conclusions, while in the Appendix we tabulate the estimated distance
modulus from the two different methods and make some few comments
on possible outliers.

\section{Two parameters correlations}

In order to be as general as possible, let us denote with $x$ a distance
independent property and let $y = k d_L^2(z)$ a given quantity with $k$ a
redshift independent constant and $d_L(z)$ the luminosity distance. Let us
suppose that we have been able to find a correlation between them such
as\,:

\begin{displaymath}
\log{y} = k + 2 \log{d_L(z)} = a + b \log{x} \ .
\end{displaymath}
Should the relation be calibrated, i.e. the parameters $(a, b)$ be known,
we can then use it to compute the distance modulus $\mu(z)$ of an object at
redshift $z$ simply as\,:

\begin{eqnarray}
\mu(z) & = & 25 + 5 \log{d_L(z)} \nonumber \\ ~ & = &
25 + (5/2) (\log{y} - k) \nonumber \\ ~ & = & 25 + (5/2) (a + b \log{x} -
k) \nonumber
\end{eqnarray}
where all the quantities are now known.

Notwithstanding its conceptual simplicity, performing such a program for
GRBs is far to be trivial. First, one has to find such a two parameters
(hereafter, 2D) correlation, i.e. decide what are the quantities $(x, y)$.
As we will see, there are different choices which seems to work
satisfactorily well, but the intrinsic scatter around the best fit line is
usually not negligible. Moreover, it is still not clear what is the
underlying physical mechanism originating the empirical 2D correlations
found. As a second more important problem, calibrating such relations is a
still open problem. Indeed, since local GRBs, i.e. GRBs with $z < 0.01$,
are not available, the only one being GRB\,980425 \cite{Galama98}, one has
typically assume an a priori cosmological model to compute $d_L(z)$ and
hence $y$ so that the calibration turns out to be model dependent thus
leading to the well known circularity problem when one aims at using GRBs
as cosmological probes. As a third problem, the calibration procedure is
performed using a variety of statistical methods so that the results from
the use of different correlations can not be straightforwardly compared.

In the following, we first briefly review the known 2D correlations and
then describe a Bayesian motivated fitting procedure to determine the
calibration parameters $(a, b)$ and the intrinsic scatter $\sigma_{int}$
once a GRB sample is available.

\subsection{2D empirical laws}

The 2D correlations we will discuss here typically relate a GRB observable
with the isotropic absolute luminosity $L$ or the collimation corrected
energy $E_{\gamma}$, with a single case correlating with the X\,-\,ray
afterglow luminosity. On the other hand, the observable properties cover a
wide range of GRBs features including both quantities related to the energy
spectrum and the light curve. In the first class, we find the peak energy
$E_{p}$, which is the energy at whith the $\nu F_{\nu}$ spectrum is
brightest, and the time lag $\tau_{lag}$, which measures the time offset
between the arrival of the low and high energy photons. To the light curve
shape are instead related the rise time $\tau_{RT}$, which is the shortest
time over which the light curve rises by half the peak flux of the burst,
and the variability $V$, which quantifies the smoothness of the light curve
itself. Somewhat different is the $L_X - T_a$ correlation, with $L_X$ the
luminosity at the time $T_a$ and $T_a$ a time scale, both referring to the
X\,-\,ray afterglow decay curve.

As yet said, neither $L$ nor $E_{\gamma}$ are directly measurable
quantities since they depend on the luminosity distance $d_L(z)$. Indeed,
it is\,:

\begin{equation}
L = 4 \pi d_L^2(z) P_{bolo} \ ,
\label{eq: lpbolo}
\end{equation}

\begin{equation}
E_{\gamma} = 4 \pi d_L^2(z) S_{bolo} F_{beam} (1 + z)^{-1}  \ .
\label{eq: egamma}
\end{equation}
Here, $P_{bolo}$ and $S_{bolo}$ are the bolometric peak flux and fluence,
respectively, while $F_{beam} = 1 - \cos{\theta_{jet}}$ is the beaming
factor with $\theta_{jet}$ the rest frame time of achromatic break in the
afterglow light curve. Note that $P_{bolo}$ and $S_{bolo}$ are computed
from the observed GRB energy spectrum $\Phi(E)$ as follows \cite{S07}\,:

\begin{equation}
P_{bolo} = P  \ {\times} \ \frac{\int_{1/(1 + z)}^{10^4/(1 + z)}{E \Phi(E) dE}}
{\int_{E_{min}}^{E_{max}}{E \Phi(E) dE}} \ ,
\label{eq: defpbolo}
\end{equation}

\begin{equation}
S_{bolo} = S \ {\times} \ \frac{\int_{1/(1 + z)}^{10^4/(1 + z)}{E \Phi(E) dE}}
{\int_{E_{min}}^{E_{max}}{E \Phi(E) dE}} \ ,
\label{eq: defsbolo}
\end{equation}
with $P$ and $S$ the observed peak energy and fluence in units of ${\rm
erg/cm^2/s}$ and ${\rm erg/cm^2}$, respectively, and $(E_{min}, E_{max})$
the detection thresholds of the observing instrument. Note that the energy
spectrum is modelled using a smoothly broken power\,-\,law \cite{Band93}
which reads\,:

\begin{equation}
\Phi(E) = \left \{
\begin{array}{ll}
A E^{\alpha} {\rm e}^{-(2 + \alpha) E/E_p} & E \le [(\alpha -
\beta)/(2 + \alpha)] E_p \\ ~ & ~ \\
B E^{\beta} & {\rm otherwise}
\end{array}
\right . \ .
\label{eq: band}
\end{equation}
While $L$ and $E_{\gamma}$ are bolometric quantities, we will also consider
a recently found 2D correlations involving only X\,-\,ray observables. In
particular, we will therefore be concerned with the luminosity in the
X\,-\,ray. As a general rule, if we denote by $L_X(t)$ and $F_X(t)$ the
luminosity and the flux at the time $t$ in the afterglow light curve, one
can write\,:

\begin{equation}
L_X(t) = 4 \pi d_L^2(z) F_X(t)
\label{eq: lx}
\end{equation}
where the flux must be $K$\,-\,corrected \cite{B01} as\,:

\begin{equation}
F_X(t) = f(t) \ {\times} \
\frac{\int_{E_{min}/(1 + z)}^{E_{max}/(1 + z)}{E \Phi_X(E)
dE}}{\int_{E_{min}}^{E_{max}}{E \Phi_X(E) dE}}
\label{eq: fx}
\end{equation}
with $(E_{min}, E_{max}) = (0.3, 10) \ {\rm keV}$ set by the instrument
bandpass and $\Phi_X(E)$ the energy spectrum in the X\,-\,ray band, well
approximated as a single power\-\,law. Finally, $f(t)$ describes the time
variation of the flux along the afterglow curve and is parametrized as
given in Willingale et al. (2007). Note that Eq.(\ref{eq: lx}) is the same
as Eq.(\ref{eq: lpbolo}) the only difference being the integration limits
at the numerator. Actually, while $L$ is the bolometric luminosity, $L_X$
refers to the X\,-\,ray one so that we integrate only over this energy
range.

As an important remark, it is worth stressing that all the 2D correlations
we will use refer to quantities evaluated in the GRB rest frame, while the
observed spectra and light curves are in the Earth orbiting satellite
frames. As such, they are affected by time dilation and redshifting which
have to be corrected for before the calibration procedure. All of the 2D
correlations available insofar are power\,-\,laws so that it is useful to
work in logarithmic units. As a general rule, we will therefore consider a
linear rule as\,:

\begin{equation}
\log{R} = a \log{Q} + b
\label{eq: linear}
\end{equation}
where $R$ is a distance dependent quantity, while $Q$ is not. Different
choices for $(R, Q)$ determine the different 2D empirical laws. Setting $R
= E_{\gamma}$ and $Q = E_p (1 + z)/300 \ {\rm keV}$ (where the factor $(1 + z)$
corrects for redshifting) gives the Ghirlanda relation \cite{G04,G06}. It
is worth noting that this is the only relation involving an energy
quantity, the other ones always involving a luminosity as $R$ quantity.
Indeed, setting $R = L$, we can find 2D correlations for $Q = E_p (1 +
z)/300 \ {\rm keV}$ \cite{S03,Yo04}, $Q = \tau_{lag} (1 + z)^{-1}/0.1 \
{\rm s}$ \cite{N00}, $Q = \tau_{RT} (1 + z)^{-1}/0.1 \ {\rm s}$ \cite{S07}
and $Q = V (1 + z)/0.02$ \cite{FRR00,R01,S07}. Note that in all these
correlations the factor $(1 + z)$ or $(1 + z)^{-1}$ corrects for time
dilation or redshifting, while the normalization is chosen in order to
minimize the correlation between the errors. Finally, setting $R =
L_X(T_a)$ and $Q  = T_a/(1 + z)$ gives the 2D correlation empirically found
for the first time in Dainotti et al. (2008) and later confirmed by
Ghisellini et al. (2008) on a semitheoretical basis and by Yamazaki (2008)
in the framework of his model proposed to explain the observed plateau in
the X\,-\,ray GRBs afterglows.

\subsection{Bayesian fitting procedure}

Eq.(\ref{eq: linear}) is a linear relation which can be fitted to a given
dataset $(Q_i, R_i)$ in order to determine the two calibration parameters
$(a, b)$. Actually, the situation is not so simple as one may naively
expect. Indeed, both the $(Q, R)$ variables are affected by measurement
uncertainties $(\sigma_Q, \sigma_R)$ which can not be neglected. Moreover,
$\sigma_Q/Q \sim \sigma_R/R$ so that the simple rule to choose as
independent variable in the fit the one with the smallest relative error
can not be applied here. Finally, the 2D correlations we are interested in
are still not definitively explained by an underlying theoretical model
determining the detailed features of the GRBs explosion and afterglow
phenomenology. Both on theoretical and observational grounds, we do expect
a certain amount of intrinsic scatter around the best fit line which has to
be taken into account and determine together with $(a, b)$ by the fitting
procedure. Different statistical recipes are available to cope with these
problems with each author having its own preferences. How the fitting
technique employed affects the final estimate of the distance modulus for a
given GRB from the different 2D correlations is not clear so that it is
highly desirable to recalibrate all of them with the same method.

\begin{table*}
\caption{Calibration parameters $(a, b)$ and intrinsic scatter $\sigma_{int}$ for the 2D
correlations $log{R} = a \log{Q} + b$ in a fiducial $\Lambda$CDM model.
Columns are as follows\,: 1. id of the correlation with the first letter
referring to the $R$ variable and the second to the $Q$ quantity; 2. number
of GRBs used; 3. maximum likelihood parameters; 4, 5, 6. median, root mean
square and Spearman rank correlation parameter with the redshift of the
best fit residuals; 7, 8. median value and $68$ and $95\%$ confidence
ranges for the parameters $(a, \sigma_{int})$.}
\begin{center}
\begin{tabular}{|c|c|c|c|c|c|c|c|c|}
\hline
Id & ${\cal{N}}$ & $(a, b, \sigma_{int})_{ML}$ & $\delta_{med}$ &
$\delta_{rms}$ & ${\cal{C}}(\delta, z)$ & $a_{-1\sigma \
-2\sigma}^{+1\sigma \ +2\sigma}$ &
$(\sigma_{int})_{-1\sigma \ -2\sigma}^{+1\sigma \ +2\sigma}$ \\ \hline
\hline

~ & ~ & ~ & ~ & ~ & ~ & ~ & ~ \\

$E_{\gamma}$\,-\,$E_p$ & 27 & (1.38, 50.56, 0.25) & 0.01 & 0.38 & -0.17 &
$1.37_{-0.26 \ -0.30}^{+0.23 \ +0.48}$ & $0.30_{-0.09 \ -0.16}^{+0.11 \
+0.28}$ \\

~ & ~ & ~ & ~ & ~ & ~ & ~ & ~ \\

$L$\,-\,$E_p$ & 64 & (1.24, 52.16, 0.45) & -0.05 & 0.51 & 0.07 &
$1.24_{-0.18 \ -0.36}^{+0.18 \ +0.36}$ & $0.48_{-0.07 \ -0.12}^{+0.07 \
+0.17}$ \\

~ & ~ & ~ & ~ & ~ & ~ & ~ & ~ \\

$L$\,-\,$\tau_{lag}$ & 38 & (-0.80, 52.28, 0.37) & 0.02 & 0.40 & 0.40 &
$-0.80_{-0.14 \ -0.30}^{+0.14 \ +0.30}$ & $0.40_{-0.07 \ -0.12}^{+0.09 \
+0.21}$ \\

~ & ~ & ~ & ~ & ~ & ~ & ~ & ~ \\

$L$\,-\,$\tau_{RT}$ & 62 & (-0.89, 52.48, 0.44) & 0.04 & 0.47 & 0.27 &
$-0.89_{-0.18 \ -0.38}^{+0.16 \ +0.31}$ & $0.46_{-0.06 \ -0.11}^{+0.07 \
+0.16}$ \\

~ & ~ & ~ & ~ & ~ & ~ & ~ & ~ \\

$L$\,-\,$V$ & 51 & (1.03, 52.49, 0.48) & -0.09 & 0.50 & 0.25 & $1.04_{-0.29
\ -0.57}^{+0.39 \ +0.60}$ & $0.51_{-0.07 \ -0.13}^{+0.09 \ +0.21}$ \\

~ & ~ & ~ & ~ & ~ & ~ & ~ & ~ \\

$L_X$\,-\,$T_a$ & 28 & (-0.58, 48.09, 0.33) & -0.12 & 0.43 & -0.21 &
$-0.58_{-0.18 \ -0.37}^{+0.18 \ +0.38}$ & $0.39_{-0.11 \ -0.20}^{+0.14
\ +0.33}$ \\

~ & ~ & ~ & ~ & ~ & ~ & ~ & ~ \\

\hline
\end{tabular}
\end{center}
\end{table*}

To this aim, in the following we will resort to a Bayesian motivated
technique \cite{Dago05} thus maximizing the likelihood function
${\cal{L}}(a, b, \sigma_{int}) = \exp{[-L(a, b, \sigma_{int})]}$ with\,:

\begin{eqnarray}
L(a, b, \sigma_{int}) & = &
\frac{1}{2} \sum{\ln{(\sigma_{int}^2 + \sigma_{R_i}^2 + a^2
\sigma_{Q_i}^2)}} \nonumber \\
~ & + & \frac{1}{2} \sum{\frac{(R_i - a Q_i - b)^2}{\sigma_{int}^2 +
\sigma_{Q_i}^2 + a^2 \sigma_{Q_i}^2}}
\label{eq: deflike}
\end{eqnarray}
where the sum is over the ${\cal{N}}$ objects in the sample. Note that,
actually, this maximization is performed in the two parameter space $(a,
\sigma_{int})$ since $b$ may be estimated analytically as\,:

\begin{equation}
b = \left [ \sum{\frac{R_i - a Q_i}{\sigma_{int}^2 + \sigma_{R_i}^2 + a^2
\sigma_{Q_i}^2}} \right ] \left [\sum{\frac{1}{\sigma_{int}^2 + \sigma_{R_i}^2 + a^2
\sigma_{Q_i}^2}} \right ]^{-1}
\label{eq: calca}
\end{equation}
so that we will not consider it anymore as a fit parameter.

It is worth noting that the Bayesian approach allows to find out what is
the most likely set of parameters within a given theory, but does not tell
us whether this model fits well or not the data. An easy way to
quantitatively estimate the goodness of the fit is obtained considering the
median and root mean square of the best fit residuals, defined as $\delta =
R_{obs} - R_{fit}$ which we will also compute for the different 2D
correlations we will consider.

The Bayesian approach used here also allows us to quantify the
uncertainties on the fit parameters. To this aim, for a given parameter
$p_i$, we first compute the marginalized likelihood ${\cal{L}}_i(p_i)$ by
integrating over the other parameter. The median value for the parameter
$p_i$ is then found by solving\,:

\begin{equation}
\int_{p_{i,min}}^{p_{i,med}}{{\cal{L}}_i(p_i) dp_i} = \frac{1}{2}
\int_{p_{i,min}}^{p_{i,max}}{{\cal{L}}_i(p_i) dp_i} \ .
\label{eq: defmaxlike}
\end{equation}
The $68\%$ ($95\%$) confidence range $(p_{i,l}, p_{i,h})$ are then found by
solving\,:

\begin{equation}
\int_{p_{i,l}}^{p_{i,med}}{{\cal{L}}_i(p_i) dp_i} = \frac{1 - \varepsilon}{2}
\int_{p_{i,min}}^{p_{i,max}}{{\cal{L}}_i(p_i) dp_i} \ ,
\label{eq: defpil}
\end{equation}

\begin{equation}
\int_{p_{i,med}}^{p_{i,h}}{{\cal{L}}_i(p_i) dp_i} = \frac{1 - \varepsilon}{2}
\int_{p_{i,min}}^{p_{i,max}}{{\cal{L}}_i(p_i) dp_i} \ ,
\label{eq: defpih}
\end{equation}
with $\varepsilon = 0.68$ (0.95) for the $68\%$ ($95\%$) range
respectively.

\section{Updating the GRB Hubble diagram}

Following the classical approach, we calibrate the six 2D correlations
described above by first estimating the distance dependent quantity
(namely, $E_{\gamma}$, $L$ or $L_X$) in a given cosmological model. In
particular, to be in agreement with the most recent results, we choose a
fiducial spatially flat concordance $\Lambda$CDM model with $\Omega_M = 1 -
\Omega_{\Lambda} = 0.291$ and $h = 0.697$ as suggested by the analysis of the
WMAP5 data \cite{WMAP5}. The sample of GRBs used is the one compiled by
Schaefer (2007) for the first five 2D correlations, while the
$L_X$\,-\,$T_a$ law is calibrated using a subsample of the Willingale et
al. (2007) catalog made out of 28 GRBs with measured redhisft,
$\log{L_X(T_a)} \ge 45$ and $1 \le \log{[T_a/(1+z)]} \le 5$ (see Dainotti
et al. 2008 for the motivation of these cuts).

\subsection{The calibration parameters}

The Bayesian fitting procedure described above then gives us the results
summarized in Table 1. It is worth stressing that, since the calibration
parameters have been obtained using the same statistical analysis for all
the 2D correlations, it is now possible to compare the relations avoiding
any possible bias introduced by the different fitting procedure.

As a first issue, one can qualitatively judge what is the best correlation,
i.e. what is the correlation giving the smaller residuals or having the
smaller intrinsic scatter. From an observational point of view, there is
not a 2D law working significantly better than the other ones since both
$\delta_{med}$  and $\delta_{rms}$ take almost the same values over the
full set, with only a modest preference for the $E_{\gamma}$\,-\,$E_p$,
$L$\,-\,$\tau_{lag}$ and $L_X$\,-\,$T_a$ correlations. On a theoretical
side, one can argue that the 2D law having the smallest intrinsic scatter
is the better motivated one since the (unknown) underlying physical
mechanism works in the same way for all the GRBs involved. Taken at face
values, the maximum likelihood estimates for $\sigma_{int}$ should argue in
favour of the $E_{\gamma}$\,-\,$E_p$ and $L_X$\,-\,$T_a$ correlations, but
this conclusion becomes meaningless when one takes into account the $68\%$
and $95\%$ confidence ranges which overlap quite well.

Finally, we caution the reader that we have implicitely assumed that the
calibration parameter do not change with the redshift notwithstanding the
fact that $z$ spans a quite large range (from $z = 0.125$ up to $z = 6.6$).
The limited number of GRBs makes it not possible to explore in detail the
validity of this usually adopted working hypothesis, but we can get a hint
by considering whether the residuals correlate with the reshift. The values
reported in Table 1 shows that only a weak and not significant correlation
exists thus arguing in favour of the starting hypothesis. This is in
agreement with what has recently been found by Basilakos \&
Perivolaropoulos (2008) which have calibrated the first five 2D
correlations in Table 1 dividing the GRBs in redshift bins. Although they
used a different fitting technique so that our results can not be directly
compared to their ones, we nonetheless agree with their conclusion that no
statistical evidence of a dependence of the $(a, b,
\sigma_{int})$ parameters on the redshift is present.

\subsection{Making up the GRBs Hubble diagram}

Once the six 2D correlations have been calibrated, we can now use them to
compute the GRBs Hubble diagram. In order to reduce the error and, in a
sense, marginalize over the possible systematic biases present in each one
of the correlations, we follow Schaefer (2007) and take a weighted average
of the distance modulus provided by each one of the six 2D laws in Table 1.
As a preliminary step, let us remember that the luminosity distance of a
GRB with redshift $z$ may be computed as\,:

\begin{equation}
d_L(z) = \left \{
\begin{array}{l}
\displaystyle{E_{\gamma} (1 + z)/4 \pi F_{beam} S_{bolo}} \\ ~ \\
\displaystyle{L/4 \pi P_{bolo}} \\ ~ \\
\displaystyle{L_X (1 + z)^{\beta_a + 2}/4 \pi F_X(T_a)} \\ ~ \\
\end{array}
\right .
\label{eq: dlval}
\end{equation}
depending on whether a correlation involving $E_{\gamma}$, $L$ or $L_X$ is
used\footnote{Note that the factor $(1 + z)^{\beta_a + 2}$ in the formula
involving $L_X$ come from the k\,-\,correction of the flux $F_X(T_a)$ with
$\beta_a$ being the slope of the power\,-\,law energy spectrum in the
X\,-\,ray band.}. The uncertainty on $d_L(z)$ is then estimated through the
naive propagation of the measurement errors on the involved quantities. In
particular, the error on the distance dependent quantity $R$ is estimated
as\,:

\begin{figure*}
\includegraphics[width=17cm]{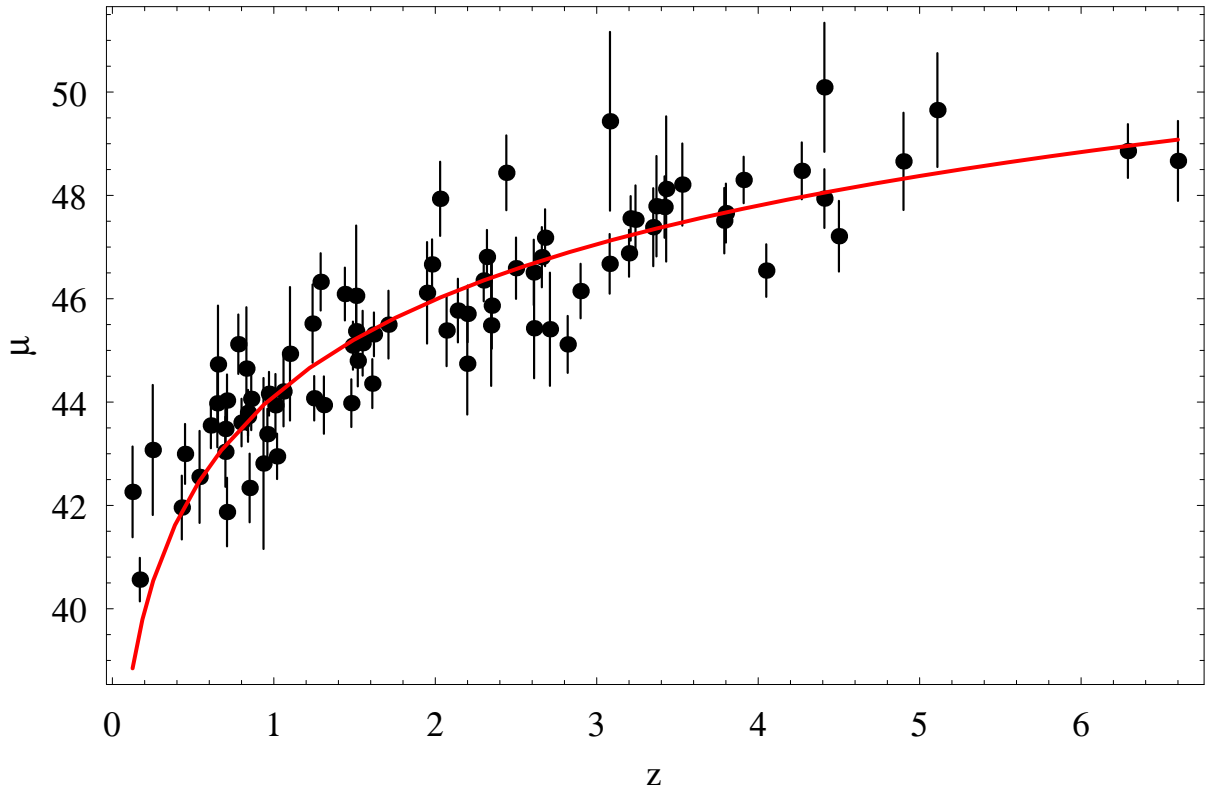}
\caption{The fiducial GRBs Hubble diagram with overplotted the
distance modulus predicted by the fiducial $\Lambda$CDM model.}
\label{fig: hdfid}
\end{figure*}

\begin{equation}
\sigma(\log{R}) = \sqrt{a^2 \sigma^2(\log{Q}) + \sigma_{int}^2}
\label{eq: siglogR}
\end{equation}
and is then added in quadrature to the other terms entering Eq.(\ref{eq:
dlval}) to get the total uncertainty. The distance modulus $\mu(z)$ is
easily obtained from its definition\,:

\begin{figure*}
\includegraphics[width=17cm]{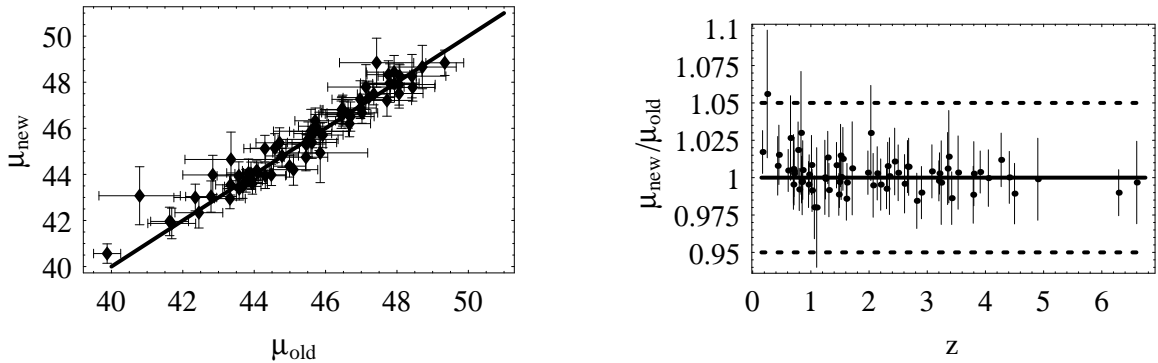}
\caption{Comparison between $\mu(z)$ for the
69 Schaefer GRBs as computed by Schaefer and by us (denoted as $\mu_{old}$
and $\mu_{new}$ respectively).}
\label{fig: oldvsnewmuval}
\end{figure*}

\begin{equation}
\mu(z) = 25 + 5 \log{d_L(z)}
\label{eq: defmu}
\end{equation}
with its uncertainty obtained again by error propagation. Following
Schaefer (2007), we finally estimate the distance modulus for the
$i$\,-\,th GRB in the sample at redshift $z_i$ as\,:

\begin{equation}
\mu(z_i) = \left [ \sum_{j}{\mu_j(z_i)/\sigma_{\mu_j}^2} \right ] \ {\times} \
\left [ \sum_{j}{1/\sigma_{\mu_j}^2} \right ]^{-1}
\label{eq: muendval}
\end{equation}
with the uncertainty given by\,:

\begin{equation}
\sigma_{\mu} = \left [ \sum_{j}{1/\sigma_{\mu_j}^2} \right ]^{-1}
\label{eq: muenderr}
\end{equation}
where the sum runs over the 2D empirical laws which can be used for the GRB
considered. Joining the Willingale et al. (2007) and Schaefer (2007)
samples and considering that 17 objects are in common, we end up with a
catalogo of 83 GRBs which we use to build the Hubble diagram plotted in
Fig.\,\ref{fig: hdfid} and tabulated in the Appendix\footnote{ASCII tables
with the Hubble diagram and all the other datasets necessary to repeat the
full analysis presented in this paper together with the {\it Mathematica}
codes used are available on request to V.F. Cardone ({\tt
winnyenodrac@gmail.com}).}. We will refer in the following to this dataset
as the {\it fiducial} GRBs Hubble diagram (hereafter, HD) since it relies
on a calibration made using a fiducial $\Lambda$CDM model to compute the
distances.

It is worth comparing our fiducial HD to the one derived by Schaefer (2007)
which we will refer to as the Schaefer HD. Actually, we have improved over
the Schaefer HD by three aspects, namely updating the $\Lambda$CDM model
parameters, using a Bayesian motivated fitting procedure and adding a
further 2D correlation. To investigate the impact of these differences, we
first consider the 69 Schaefer GRBs and compute the distance modulus with
our new calibration, but not including the values coming from the
$L_X$\,-\,$T_a$ correlation. We then plot the in left panel of
Fig.\,\ref{fig: oldvsnewmuval} the old vs the new estimates of the distance
modulus. As can be seen, there is a clear one\,-\,to\,-\,one correspondence
with slope perfectly compatible with 1 and no systematic offset. This can
also better appreciated looking at the right panel showing that
$\mu_{new}/\mu_{old}$ is close to 1 within the $5\%$ with no correlation
with the redshift. To be more quantitative, averaging over the 69 GRBs, we
find $\mu_{new}/\mu_{old} = 1.00 {\pm} 0.01$ with an root mean square value
$(\mu_{new}/\mu_{old})_{rms} = 1.002$. We may therefore safely conclude
that the new calibration procedure have not affected the results.

\begin{figure*}
\includegraphics[width=17cm]{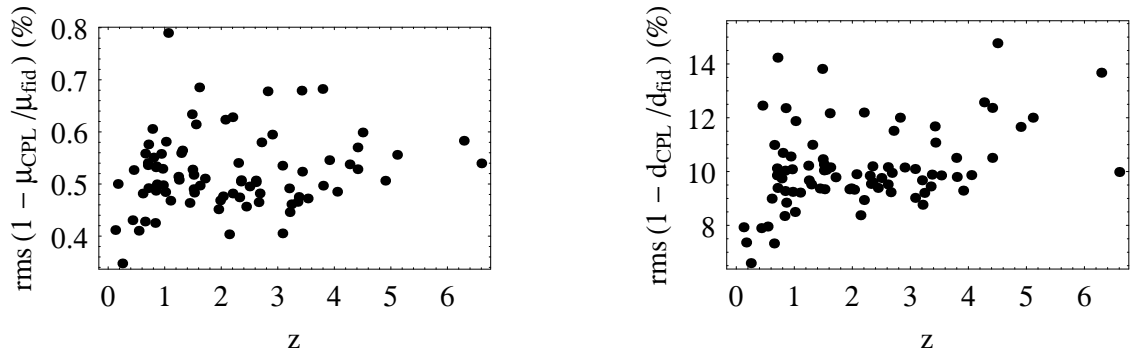}
\caption{Root mean square of the percentage deviation of the distance modulus $\mu(z)$
(left) and the luminosity distance $d_L(z)$ from the fiducial $\Lambda$CDM
values caused by changing the $(w_0, w_a)$ parameters for $\Omega_M
= 0.30$. Similar results are obtained for other $\Omega_M$ values.}
\label{fig: devmudlcpl}
\end{figure*}

A cautionary remark on the errors is in order here. As explained above, the
uncertainty on $\mu$ comes from the propagation of the errors starting from
the ones on the distance dependent quantities $E_{\gamma}$, $L$ or $L_X$.
Comparing Eq.(\ref{eq: siglogR}) with the similar formulae in Schaefer
(2007), one can see that we have not included in the propagation the errors
on the calibration coefficient $(a, b)$. This is a direct consequence of
the different statistical phylosophy underlying the fitting procedure
employed (Bayesian maximum likelihood here vs the frequentist $\chi^2$
minimization in Schaefer). We have however checked whether this bias
somewhat the error estimates. Averaging over the 69 GRBs, we find
$\sigma_{new}/\sigma_{old} = 1.01 {\pm} 0.06$ with a root mean square value
$(\sigma_{new}/\sigma_{old})_{rms} = 1.01$ and no correlation with the
redshift. These numbers make us then confident that no under or
overestimation of the errors is introduced by our different approach.

Having added the new $L_X$\,-\,$T_a$ correlation makes it possible to infer
the distance modulus for further 14 objects thus leading to 83 the number 
of GRBs in the fiducial HD, i.e. $20\%$ more data. Moreover, this also makes it 
possible to reduce the error thanks to the weighted averaging procedure
employed. Indeed, in the old Schaefer HD, the numbers of GRBs with $\mu(z)$
estimated from 1 to 5 correlations are respectively (3, 12, 14, 27, 13). On
the contrary, these numbers now read (17, 7, 16, 24, 16, 3) with 3 GRBs
having $\mu$ computed from all the six 2D correlations. We do expect that
this lead to a decrease of the uncertainty on $\mu(z)$. Averaging over the
83 objects, we indeed find $\sigma_{\mu}/\sigma_{Sch} = 0.86 {\pm} 0.09$ with
$(\sigma_{\mu}/\sigma_{Sch})_{rms} = 0.87$, i.e. there is on average a
significative $14\%$ reduction.

Summarizing, the detailed analysis performed make us confident that the
fiducial GRB HD plotted in Fig.\,\ref{fig: hdfid} and tabulated in the
Appendix represents a noteworthy step forward for using GRBs as
cosmological tools.

\subsection{Varying the cosmological model}

The fiducial HD has been obtained after calibrating the six 2D correlations
in Table 1 assuming a $\Lambda$CDM concordance cosmology. Although this
model is in agreement with most of the available astrohysical data, there
is still not a clear consensus on the value of its parameters $(\Omega_M,
h)$. Moreover, although not preferred by the data, a varying equation of
state (EoS) for the dark energy fluid is still a viable (and theoretically
better motivated) option. A large class of dark energy models predict a
depedence of the EoS on the scale factor $a$ which is well fitted by the
Chevallier\,-\,Polarski\,-\,Linder (CPL) ansatz \cite{CP01,L03}\,:

\begin{equation}
w(a) = w_0 + w_a (1 - a) = w_0 + w_a z/(1 + z) \ .
\label{eq: cpleos}
\end{equation}
The luminosity distance is then computed as\,:

\begin{equation}
d_L(z) = \frac{c}{H_0} (1 + z) \int_{0}^{z}{\frac{dz'}{E(z')}}
\label{eq: cpldl}
\end{equation}
with $E(z) = H(z)/H_0$ given by\,:

\begin{equation}
E^2(z) = \Omega_M (1 + z)^3 + \Omega_X (1 + z)^{3(1 + w_0 + w_a)} {\rm
e}^{-\frac{3 w_a z}{1 + z}}
\label{eq: ecpl}
\end{equation}
where $\Omega_X = 1 - \Omega_M$ because of the flatness assumption.

In order to quantify the impact of varying the EoS parameters $(w_0, w_a)$
and the matter density $\Omega_M$, we have recalibrated all the six 2D
correlations on a regular grid in $(w_0, w_a)$ for five different values of
$\Omega_M$. Namely, we consider CPL models with $-1.3 \le w_0 \le -0.3$ and
$-1.0 \le w_a \le 1.0$ for $\Omega_M$ from 0.20 to 0.40 in steps of 0.05.
These ranges are larger than what is allowed by the data, but have been
chosen in order to take into account also models very different from the
fiducial one. For each point in the grid, we repeat all the steps described
in the previous subsection to get the distance modulus to each GRB in the
sample. We then collect these values and evaluate, for each GRB, the root
mean square of the percentage deviation from the fiducial $\mu$ value. The
results are plotted in the left panel of Fig.\,\ref{fig: devmudlcpl}. Note
that in this way we are performing a sort of average of the absolute
percentage deviation so that, for instance, we can say that the distance
modulus to a given GRB varies of order the number in the plot when $(w_0,
w_a)$ span the range considered above. Although averaging over different
models is not motivated, we have checked that this simple procedure gives
us the correct order of magnitude of the effect of changing the EoS and the
matter content on the final Hubble diagram.

Looking at the left panel of Fig.\,\ref{fig: devmudlcpl}, it is clear that
the uncertainty on the background cosmological model to use in the
calibration procedure is quite modest. Indeed, the distance modulus may be
under or overestimated by a modest $0.5\%$ with values never larger than
$1\%$. This is is the same order of magnitude and often smaller than the
measurement uncertainty on $\mu(z)$ so that one can argue that our fiducial
GRBs HD may be used even if the underlying cosmological model is different
from the $\Lambda$CDM one we have assumed. Actually, this result is partly
due to the use of the distance modulus rather than the luminosity distance
directly. Indeed, as well known, a logarithmic scale always reduce the
errors, an effect which is at work also in this case as can be understood
looking at the right panel of Fig.\,\ref{fig: devmudlcpl}. Here, we plot
the same quantity as the left panel, but referring to the luminosity
distance $d_L(z)$. The rms percentage deviation now increases to $\sim
10\%$ with values as large as $15\%$ so that the impact of cosmology is more
apparent. However, it is worth stressing that it is common practice to use
the $\mu(z)$ rather than the $d_L(z)$ data when constraining cosmological
models so that we can give off any further consideration of the
reconstructed luminosity distance.

\begin{figure}
\includegraphics[width=8.5cm]{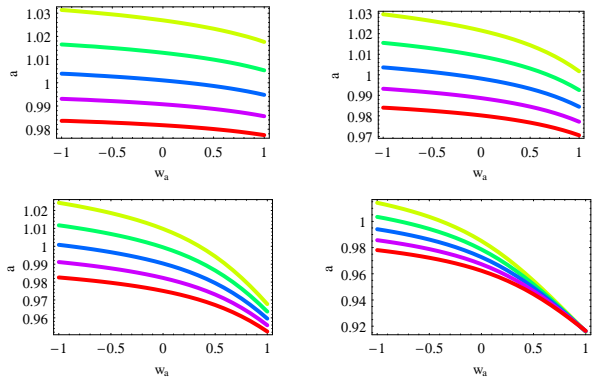}
\caption{The dependence of the $a$ coefficient of the $E_{\gamma}$\,-\,$E_p$ correlation
normalized to the fiducial value on the $w_a$ parameter for different
values of $w_0$ (namely, $w_0 = -1.3, -1.0, -0.7, -0.4$ from top left to
bottom right) and $\Omega_M$ (from 0.20 to 0.40 in steps of 0.05 from
yellow higher to red lower curve).}
\label{fig: apar}
\end{figure}

\begin{figure}
\includegraphics[width=8.5cm]{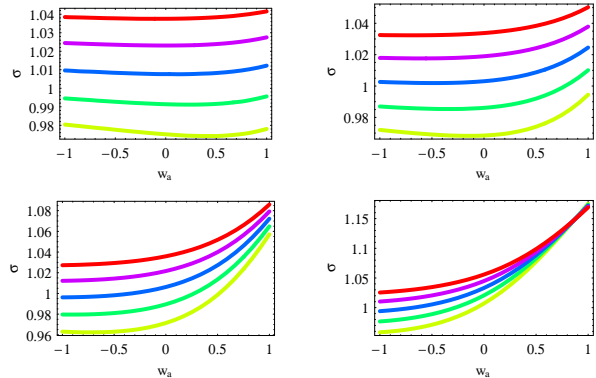}
\caption{Same as Fig.\,\ref{fig: apar}, but for the intrinsic scatter $\sigma_{int}$. Note that
$\Omega_M$ takes the same values as before, but now increases from the
yellow lower to the red higher curve in each panel.}
\label{fig: sint}
\end{figure}

It is somewhat surprising that changing the cosmological model used for the
calibration has such a small impact on the estimated $\mu(z)$. However, it
is easy to trace back the origin of this unexpcted result. Figs.\,\ref{fig:
apar} and \ref{fig: sint} show the dependence on the $(\Omega_M, w_0, w_a)$
parameters of the slope coefficient $a$ and the intrinsic scatter
$\sigma_{int}$ (normalized to their fiducial values) of the
$E_{\gamma}$\,-\,$E_p$ correlation. Although there are clear trends, both
$a$ and $\sigma_{int}$ change of order few $\%$ for the cosmological
parameters spanning their quite conservative ranges. This is also the case
for the other 2D correlations we have used so that we can deem this as a
general result. This result then suggests that the calibration is in
practice model independent, at least within the precision allowed by the
present measurement uncertainties. As a consequence, the final Hubble
diagram is almost unaffected by the choice of the cosmological model
provided this can be well described within the framework of the CPL
parameterization.

\section{Local regression calibration}

Although the above analysis has shown that the choice of the underlying
cosmological model has only a modest impact on the final estimate of the
distance modulus, it is worth stressing that such a result is far to be
definitive. First, we have considered only a single (although large) class
of dark energy models. Moreover, one can argue that future data will be
affected by smaller measurement uncertainties so that the impact of
cosmology may turn out to be more important. It is therefore of vital
importance to look for a method to calibrate the 2D correlations through a
method that do not rely on the assumption of any cosmological model.

Due to the lack of a set of low redshift GRBs at $z < 0.1$ which are
cosmology independent, the so called {\it circularity problem} arise\,: in
order to use the GRBs as cosmological tools, one has to calibrate the 2D
correlations, but a cosmological model has to be assumed for the
calibration. In principle, such a problem could be avoided in two ways.
First, a solid theoretical model must be found in order to physically
motivate the observed 2D correlations thus setting their calibration
parameters. Unfortunately, finding such a model ramains an ambitious, but
still unsuccessful task. Alternatively, it has been proposed to avoid the
need for any distance determination by using a sufficiently large sample of
GRBs within a small redshift bin centred at a whatever $z$ \cite{G06,LZ06}.
However, this method might be unrealistic, since the current sample of
observed GRBs is not large enough. It has recently been suggested that such
a model independent calibrations may be carried out using SNeIa as distance
indicator based on the naive observations that a GRBs at redshift $z$ must
have the same distance modulus of SNeIa having the same redshift.
Interpolating therefore the SNeIa Hubble diagram gives therefore the value
of $\mu(z)$ for a subset of the GRBs sample with $z \le 1.4$ which can then
be used to calibrate the 2D correlations \cite{K08,L08,WZ08}. Assuming that
this calibration is redshift independent, one can then build up the Hubble
diagram at higher redshifts using the calibrated correlations for the
remaining GRBs in the sample. We present here a similar approach still
using the SNeIa as secondary distance indicator, but avoiding the need for
interpolating the sparse dataset through the use of the local regression
method we briefly describe in the following.

\subsection{Local regression on SNeIa}

As a general rule, any interpolation procedure may be seen as an attempt of
recovering the true underlying curve from a sparse dataset. In a sense,
interpolating methods try to extract the order of a smooth model from the
disorder of the data originating from that same model. Interpolation may
therefore be seen as a smoothing procedure so that one can naturally resort
to the well tested strategies developed in this field. A particularly
efficient method is known as {\it local regression} (see Loader 1999 and
references therein).

\begin{table*}
\caption{Calibration parameters $(a, b)$ and intrinsic scatter
$\sigma_{int}$ for the 2D correlations $log{R} = a \log{Q} + b$
for GRBs with $z \le 1.4$ and distance modulus estimated from the
local regression on SNeIa. Columns as in Table 1.}
\begin{center}
\begin{tabular}{|c|c|c|c|c|c|c|c|c|}
\hline Id & ${\cal{N}}$ & $(a, b, \sigma_{int})_{ML}$ &
$\delta_{med}$ & $\delta_{rms}$ & ${\cal{C}}(\delta, z)$ &
$a_{-1\sigma \ -2\sigma}^{+1\sigma \ +2\sigma}$ &
$(\sigma_{int})_{-1\sigma \ -2\sigma}^{+1\sigma \ +2\sigma}$ \\
\hline \hline

~ & ~ & ~ & ~ & ~ & ~ & ~ & ~ \\

$E_{\gamma}$\,-\,$E_p$ & 12 & (1.53, 50.59, 0.12) & 0.10 & 0.22 &
0.22 & $1.54_{-0.27 \ -0.66}^{+0.30 \ +0.74}$ & $0.19_{-0.11 \
-0.17}^{+0.17 \ +0.43}$ \\

~ & ~ & ~ & ~ & ~ & ~ & ~ & ~ \\

$L$\,-\,$E_p$ & 25 & (1.40, 52.20, 0.40) & 0.06 & 0.45 & 0.12 &
$1.40_{-0.24 \ -0.51}^{+0.25 \ +0.53}$ & $0.46_{-0.10 \
-0.18}^{+0.14 \ +0.34}$ \\

~ & ~ & ~ & ~ & ~ & ~ & ~ & ~ \\

$L$\,-\,$\tau_{lag}$ & 13 & (-0.89, 52.07, 0.32) & 0.11 & 0.35 &
0.33 & $-0.90_{-0.27 \ -0.64}^{+0.27 \ +0.61}$ & $0.42_{-0.13 \
-0.22}^{+0.23 \ +0.61}$ \\

~ & ~ & ~ & ~ & ~ & ~ & ~ & ~ \\

$L$\,-\,$\tau_{RT}$ & 24 & (-0.97, 52.40, 0.44) & 0.04 & 0.47 &
0.33 & $-0.97_{-0.27 \ -0.57}^{+0.27 \ +0.57}$ & $0.50_{-0.11 \
-0.18}^{+0.15 \ +0.36}$ \\

~ & ~ & ~ & ~ & ~ & ~ & ~ & ~ \\

$L$\,-\,$V$ & 19 & (1.14, 52.33, 0.54) & -0.04 & 0.56 & 0.24 &
$1.19_{-0.61 \ -1.39}^{+0.67 \ +1.47}$ & $0.64_{-0.15 \ -0.25}^{+0.22 \ +0.57}$ \\

~ & ~ & ~ & ~ & ~ & ~ & ~ & ~ \\

\hline
\end{tabular}
\end{center}
\end{table*}

Originally proposed by Cleveland (1979) and further developed by Cleveland
and Devlin (1988), the local regression technique combines much of the
simplicity of linear least squares regression with the flexibility of
nonlinear regression. The basic idea relies on fitting simple models to
localized subsets of the data to build up a function that describes the
deterministic part of the variation in the data, point by point. Actually,
one is not required to specify a global function of any form to fit a model
to the data so that there is no ambiguity in the choice of the
interpolating function. Indeed, at each point, a low degree polynomial is
fit to a subset of the data containing only those points which are nearest
to the point whose response is being estimated. The polynomial is fit using
weighted least squares with a weight function which quickly decreases with
the distance from the point where the model has to be recovered.

We apply local regression to estimate the distance modulus $\mu(z)$ from
the most updated SNeIa sample. Referred to as Union \cite{Union}, this
compilation includes recent large samples from SNLS \cite{SNLS} and ESSENCE
\cite{ESSENCE} surveys, older data sets and the recently extended data set
of distant SNeIa observed with HST, all of which have been homogenously
reanalyzed with the same lightcurve fitter. After selection cuts and
outliers removal, the final sample contains 307 SNeIa spanning the range
$0.015 \le z \le 1.55$. We use this large dataset as input to the local
regression estimate of $\mu(z)$ following the steps schematically sketched
below.

\begin{enumerate}

\item{Set a redshift $z$ where $\mu(z)$ have to recovered. \\}

\item{Order the SNeIa according to increasing value of $|z - z_i|$ and 
select the first $n = \alpha {\cal{N}}_{SNeIa}$ with $\alpha$ a user 
selected value and ${\cal{N}}_{SNeIa}$ the total number of SNeIa. \\}

\item{Define the weight function\,:

\begin{equation}
W(u) = \left \{
\begin{array}{ll}
(1 - |u|^3)^3 & |u| \le 1 \\ ~ & ~ \\ 0 & |u| \ge 1
\end{array}
\right .
\label{eq: wdef}
\end{equation}
where $u = |z - z_i|/\Delta$ and $\Delta$ the maximum value of the $|z -
z_i|$ over the subset chosen before. \\ }

\item{Fit a first order polynomial to the data selected at step (ii) weighting 
each SNeIa with the corresponding value of the function $W(u)$ and take the 
zeroth order term as best estimate of $\mu(z)$. \\ }

\item{Estimate the error on $\mu(z)$ as the root mean square of the weighted 
residuals with respect to the best fit zeroth order term. \\ }

\end{enumerate}
It is worth stressing that both the choice of the weight function and the order 
of the fitting polynomial are somewhat arbitrary. Similarly, the value of $\alpha$ 
to be used must not be too small in order to make up a statistical valuable sample, 
but also not too large to prevent the use of a low order polynomial. In order to find 
what is the better value for $\alpha$ and simultaneously check that our choices for the 
polynomial degree and weight function do not affect the reconstruction, we have performed 
an extensive set of simulations. To this aim, we use the CPL EoS and set the model parameters 
$(\Omega_M, w_0, w_a, h)$ randomly extracting from the conservative ranges 
$0.15 \le \Omega_M \le 0.45$, $-1.5 \le w_0 \le -0.5$, $-2.0 \le w_a \le 2.0$, 
$0.60 \le h \le 0.80$. For each redshift value in the Union sample, we extract $\mu(z_i)$ from 
a Gaussian distribution centred on the theoretically predicted value and with a standard 
deviation $\sigma = 0.15$, in agreement with the estimated intrinsic scatter of the SNeIa 
absolute magnitude. To this value, we then attach an error in such a way that the relative 
uncertainty is the same as the corresponding point in the Union sample thus ending up with 
a mock catalogue having the same redshift and error distribution of the actual one. This mock 
catalogue is used as input to the routine sketched above and finally the reconstructed 
$\mu(z)$ value for each point in the catalog are compared to the input one.

Defining the percentage deviation $\delta \mu/\mu = 1 -
\mu_{lr}(z)/\mu_{CPL}(z)$ with $\mu_{lr}$ and $\mu_{CPL}$ the local
regression estimate and the input CPL values respectively and averaging
over 250 realizations, we find that the above routine with the choice
$\alpha = 0.025$ gives $(\delta \mu/\mu)_{rms} \simeq 0.35\%$ with $|\delta
\mu/\mu| \le 1\%$ independent on the redshift $z$. Moreover, there is no
correlation at all of $\delta \mu/\mu$ with any of the cosmological
parameters. We therefore safely conclude that the local regression method
allows to correctly recover the underlying distance modulus at a whatever
redshift $z$ from the Union SNeIa sample whichever is the background
cosmological model.

\begin{figure*}
\includegraphics[width=17cm]{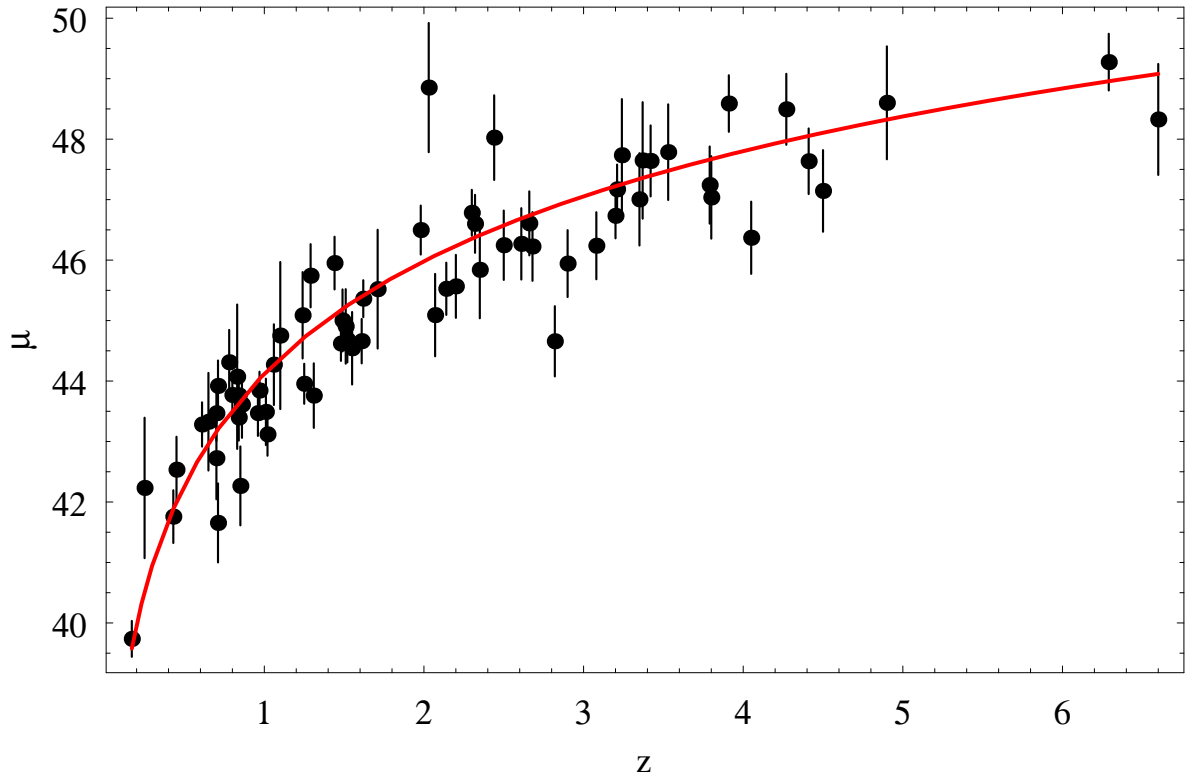}
\caption{The calibrated Hubble diagram with overplotted the
distance modulus predicted by the fiducial $\Lambda$CDM model.}
\label{fig: hdcal}
\end{figure*}

\subsection{Model independent calibration}

Having found an efficient way of estimating the distance modulus
at redshift $z$ in a model independent way, we can now calibrate
the GRBs 2D correlation using a different approach. First, we
consider only GRBs with $z \le 1.4$ in order to cover the same
redshift range spanned by the SNeIa data. As an example, let us
consider the $L$\,-\,$E_p$ correlation. For each GRBs with
measured $(z, P_{bolo})$, one can write\,:

\begin{eqnarray}
\log{L} & = & \log{(4 \pi P_{bolo})} + 2 \log{d_L(z)} \nonumber \\
~ & = & \log{(4 \pi P_{bolo})} + (2/5) \left [ \mu_{lr}(z) - 25 \right ] \nonumber \ ,
\end{eqnarray}
where we have used Eq.(\ref{eq: defmu}) and approximated the true
$\mu(z)$ with its local regression estimate $\mu_{lr}(z)$ without
any significative loss of precision. The uncertainty on $\log{L}$
is then evaluated as usual by propagation of errors so that we
finally end with a subset of GRBs with model independent estimates
of $\log{L}$ that can be used as input to the Bayesian fitting
procedure described in Sect. 2.2. Carrying on this exercise for
the first five 2D correlations in Table 1, we get the results
summarized in Table 2. Note that we cannot apply this method to
the $L_X$\,-\,$T_a$ correlation since there are only 3 GRBs with
$z \le 1.4$ after the selection cuts on $\log{[T_a/(1 + z)]}$ and
$\log{L_X}$ so that the fitting fails.

Comparing the values in Tables 1 and 2, one should get some
insight on the dependence of the calibration coefficients on the
redshift. Taken at face values, it is apparent that fitting only
$z \le 1.4$ GRBs steepens all the 2D correlations, while the
intrinsic scatter is almost unchanged in all cases but the
$E_{\gamma}$\,-\,$E_p$ correlation. While this could naively be
interpreted as an evidence for evolution with redshift of at least
the slope, such a result loses its statistical significance when
one looks at the marginalized constraints. Indeed, the $68$ and
$95\%$ confidence ranges are quite larger than in the previous
case so that inferring any constraints on the evolution of $a$ is
meaningless. Moreover, the present calibration is model
independent, while the results in Table 1 have been obtained
assuming a fiducial $\Lambda$CDM model. Although the discussion in
Sect. 3.3 and Figs.\,\ref{fig: apar} and \ref{fig: sint} suggest
that this should not be a serious problem, it is nevertheless
worth being cautious before drawing any definitive conclusion.

\begin{figure*}
\includegraphics[width=17cm]{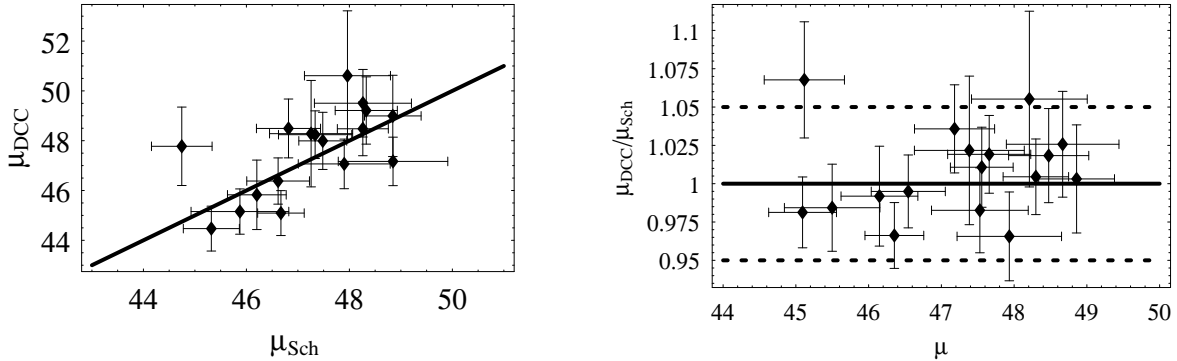}
\caption{Comparison between $\mu(z)$ for the
17 GRBs common to the Schaefer and Willingale et al. samples.}
\label{fig: dccvssch}
\end{figure*}

\subsection{The calibrated Hubble diagram}

We can now use the calibration parameter derived above to infer
the value of $\log{R}$ (with $R = E_{\gamma}$ or $L$) to construct
a new GRBs Hubble diagram following the same method used in Sect.
3.2. The result is what we refer to as the {\it calibrated} GRBs
HD. Plotted in Fig.\,\ref{fig: hdcal} and tabulated in the
Appendix, this HD now contains only the 69 Schaefer GRBs since we
have not used the $L_X$\,-\,$T_a$ correlation being impossible to
calibrate it with the local regression based method.

A visual comparison of Figs.\,\ref{fig: hdfid} and \ref{fig:
hdcal} shows that, although containing a different number of
objects (83 vs 69), the fiducial and calibrated HDs are in well
agreement. This is not unexpected since the fiducial $\Lambda$CDM
model fits quite well the SNeIa data so that the predicted values
for $\mu(z)$ are in very good agreement with the local regression
estimates based on the same SNeIa. This qualitative agreement is
also confirmed by considering the numbers of GRBs that are more
than $2 \sigma$ away from the theoretical fiducial $\Lambda$CDM
HD. For the fiducial HD, we find 14 GRBs deviating more than $2
\sigma$, i.e. $17\%$ of the sample, while this number reduces to
10, i.e. $14\%$, for the calibrated HD dataset. Moreover, the GRBs
are almost the same\footnote{The likely outliers GRBs for the
fiducial dataset are\,: 990123, 991208, 991216, 000210, 010222,
020903, 030329, 030528, 050126, 050406, 060108, 060206, 060614.
For the calibrated dataset, the list contains\,: 990123, 990506,
991208, 991216, 000210, 020813, 050406, 050603, 060108, 060206.
There are thus 7 GRBs in common within the two lists. Note that
050603 is $3 \sigma$ away in both datasets, while 060614 deviates
more than $3 \sigma$ from the fiducial HD only. We also stress
that only three GRBs (namely 050603, 060108, 060206) have $\mu(z)$
estimated also resorting to the $L_X$\,-\,$T_a$ correlation so
that the use of this new empirical law can not be considered as a
possible bias.} thus arguing that there is some problem with the
data on this GRBs rather than with the procedure used to make up
both the fiducial and the calibrated HDs. However, since we do not
actually know whether the adopted fiducial $\Lambda$CDM model is
indeed the correct one, we prefer to not reject these points in
absence of a clear alternative motivation.

To quantitatively compare the fiducial and calibrated GRBs HDs, we
have used a linear regression (not taking into account the errors
on the two variables) to determine the slope of the
$\mu_{fid}$\,-\,$\mu_{cal}$ relation, with $\mu_{fid}$ and
$\mu_{cal}$ the distance modulus in the fiducial and calibrated
HDs respectively. We find\,:

\begin{displaymath}
\mu_{fid} = (1.02 {\pm} 0.02) \mu_{cal} + (-1.28 {\pm} 0.94)
\end{displaymath}
so that the slope is consistent with $1$. It is worth stressing
that the zeropoint of this relation does not signal a systematic
offset, but is only a consequence of the slope being slightly
different from 1. Indeed, defining $\Delta \mu = \mu_{fid} -
\mu_{cal}$ and the averaging over the 69 GRBs in common, we find
$\langle \Delta \mu \rangle = 0.20$ which is consistent with a
null value if one takes into account the statistical errors on
$\mu_{fid}$ and $\mu_{cal}$.

Motivated by this successful comparison, we therefore conclude
that the local regression based calibration has not biased anyway 
the derivation of the HD so that both the fiducial and the calibrated 
GRBs HDs can hence be used as valid tools to constrain cosmological models.

\section{Possible systematics and biases}

While the previous analysis has shown that the calibrated and fiducial HDs
are consistent with each other, such a test does not tell us nothing about
possible systematic error related to the derivation of the underlying 
GRBs 2D correlations. Although a detailed analysis of the different systematic
errors that can affect these relations is outside our aims here, we nevertheless
investigate some related issues in order to qualitatively infer whether they
can bias or not the derived distance moduli.

\subsection{The $L_X$\,-\,$T_a$ correlation}

The use of the $L_X$\,-\,$T_a$ correlation have allowed us to increase the
GRBs sample from 69 to 83 objects\footnote{Note that, here, we only refer to
the fiducial GRBs HD since, as explained before, the local regression calibration
of the $L_X$\,-\,$T_a$ correlation is not possible so that this 2D law has not 
been used in the derivation of the calibrated HD.}. However, since this is the 
first time such a correlation is used to build a GRBs HD, it is worth whehter it 
is in agreement with the previous results. To this end, we have compared the estimated
$\mu(z)$ for the 17 GRBs common to the Schaefer and Willingale et al.
samples. This is shown in the left panel of Fig.\,\ref{fig: dccvssch} where
we plot $\mu_{DCC}$ vs $\mu_{Sch}$, i.e. the distance moduli as
computed from the $L_X$\,-\,$T_a$ and the other 2D correlations,
respectively. Although with a large scatter, the data are consistent with a
linear relation with slope 1, as it is also demonstrated by the right panel
where $\mu_{DCC}/\mu_{Sch}$ is next to 1 within $5\%$ for all the points,
but two (the only marginally deviant cases being GRB\,050603 and
GRB\,060115). Averaging over the 17 GRBs, we find $\mu_{DCC}/\mu_{Sch} =
1.01 {\pm} 0.03$ with $(\mu_{DCC}/\mu_{Sch})_{rms} = 1.01$ and no correlation
with redshift. As a further test, a linear regression (not weighting the points
with their errors) gives\,:

\begin{displaymath}
\mu_{DCC} = (0.9 {\pm} 0.3) \mu_{Sch} + (5.5 {\pm} 13.5)
\end{displaymath}
so that the slope is indeed consistent with 1. The small number of objects
and the large intrinsic scatter makes the errors on the coefficient extremely large
so that the offset detected is fully devoided of any stastistical meaning. 
However, that such an offset is not present is also suggested by noting that 
$\langle \mu_{Sch} - \mu_{DCC} \rangle = -0.35 \pm 1.3$ so that indeed 
not statistically meaningful systematic offset is detected. Repeating the same 
linear fit but excluding the two points with $\mu_{DCC}/\mu_{Sch} > 1.05$ gives\,:

\begin{displaymath}
\mu_{DCC} = (1.2 {\pm} 0.3) \mu_{Sch} + (-8.6 {\pm} 12.2) \ ,
\end{displaymath}

\begin{displaymath}
\langle \mu_{Sch} - \mu_{DCC} \rangle = -0.02 \pm 1.03 \ , 
\end{displaymath}
thus showing that they are not biasing the comparison.

We therefore conclude that the use of the $L_X$\,-\,$T_a$
correlation does not introduce any bias and hence can be invoked to
increase the number of GRBs in the fiducial HD. On the other hand, 
should future data disprove the $L_X$\,-\,$T_a$ correlation, one should 
reject from the fiducial HD the 14 GRBs having $\mu$ determined by 
this empirical law only so that one gets back the Schaefer Hubble diagram. 

\begin{figure*}
\includegraphics[width=17cm]{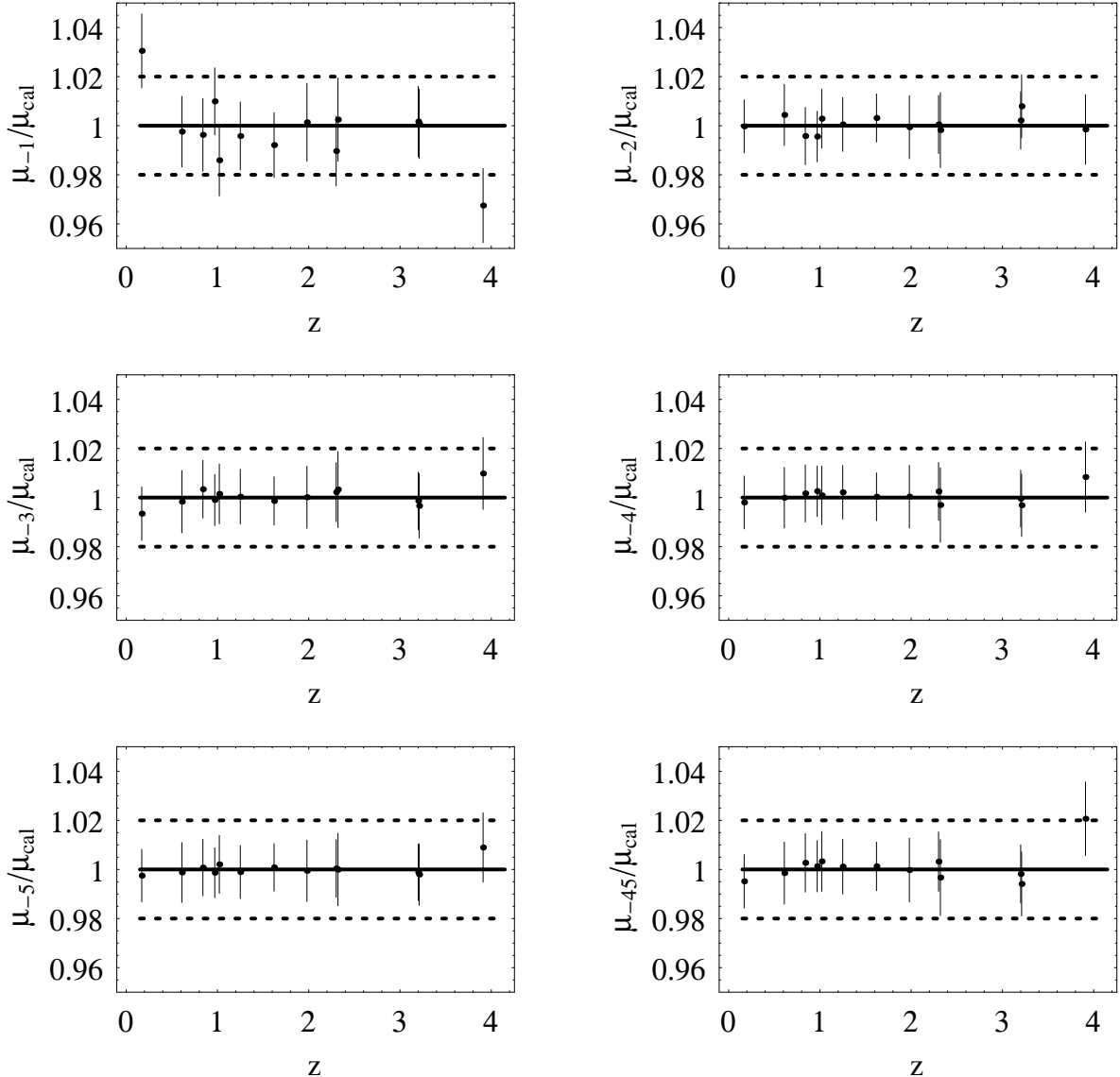}
\caption{The ratio among the $\mu_{-i}$ and $\mu_{cal}$ values with 
$\mu_{-i}$ defined in the text and $\mu_{cal}$ the calibrated distance modulus.}
\label{fig: testnocorr}
\end{figure*}

\subsection{How many correlations ?}

Although a lot of attention has been devoted recently to look for
empirical correlations among GRB properties, it is still not clear
whether the 2D laws we have used here are actually all real correlations
or a consequence of different observational selection effects. Moreover,
it is worth noting that five of them involve a luminosity measurement (the 
bolometric one, $L$, or the X\,-,ray one, $L_X$) and three of them
rely on a time measurement (being either $\tau_{lag}$, $\tau_{RT}$ or
$T_a$) so that one should also investigate whether they are actually
independent relations. Investigating this issue in detail is outside
our aims here, but we can nevertheless estimate what should be the impact
of giving off one of the relations. To this end, we consider the 
calibrated\footnote{We prefer to use the calibrated rather than the fiducial
HD in order to minimize the impact of choosing a cosmological model to determine
the parameters of each correlation.} HD and select the 13 GRBs having $\mu$ 
estimated by all the canonical five correlations. We then denote with $\mu_{-i}$
the value of $\mu$ obtained by excluding the correlation $i$ with $i = 1, \ldots, 5$
for the $E_{\gamma}$\,-\,$E_p$, $L$\,-\,$E_p$, $L$\,-\,$\tau_{lag}$, $L$\,-\,$\tau_{RT}$
and $L$\,-\,$V$ respectively. We also denote by $\mu_{-45}$ the value obtained
by excluding the $L$\,-\,$\tau_{RT}$ and $L$\,-\,$V$ correlations since they are
those with the largest intrinsic scatter. Fig.\,\ref{fig: testnocorr} shows that
the ratios $\mu_{-i}/\mu_{cal}$ (with $\mu_{cal}$ the value reported in the HD) 
is very close to 1, with only two points deviating more than $4\%$ from unity. 
Moreover, there is no apparent differences between the different panels and no
trend with the redshift or the distance modulus. Although a more quantitative 
analysis with a larger number of objects is needed to draw a definitive
answer, we may, however, argue from this preliminary analysis that the exclusion
of one or two correlations does not alter the results. Put in other words, should
one of the canonical 2D laws employed to get the calibrated HD be dismissed because
of future evidences, we should not be afraid of any bias in the derived distance moduli.

\subsection{The high $z$ GRBs}

Our sample includes 3 GRBs with $z > 5$, namely 060522
at $z = 5.11$, 050904 at $z = 6.29$ and 060116 at $z = 6.60$. 
As is apparent from Fig.\,\ref{fig: hdfid}, there is a clear 
gap between the two $z > 6$ GRBs and the rest of the sample so 
that one could wonder whether these two objects share the same 
properties of the other GRBs. 

Actually, even if extreme in redshift, both 050904 and 060522 follow
quite well the correlations they are involved in. In particular, 050904
does not deviate from five 2D laws, while the observed properties of 060522 
make this object in good agreement with three 2D correlations. As such,
we do not expect they bias anyway the fiducial HD. Moreover, being at such
a high redshift, they are not used in the calibration of the 2D empirical laws
based on the local regression analysis. Nevertheless, their distance moduli 
in the fiducial and calibrated HDs are in very good agreement. We therefore
conclude that, notwithstanding the doubts on the validity of the 2D 
correlations at very high $z$, the inclusion of these two GRBs does not
bias anyway the results.

\subsection{Selection effects}

Since the first works on the existence of correlations among
GRBs properties allowing to use these objects as standardizeable
candles, serious doubts were raised on the actual validity of 
such relations. Although being very luminous, the detection of GRBs 
is a complicated task so that different biases may favour the 
detection of a particular class of GRBs. Should this be indeed the case,
one could argue that the 2D empirical laws are not universal or, what is
worst, they only come out as a consequence of a detection bias (see, e.g., 
Butler et al. 2007 and Shahmoradi \& Nemiroff 2009). 

One possible example is represented by the shape of the GRB spectrum
in the high energy region. GRBs more fluent at high energies (in their
rest frame) at high redshift will be more likely to trigger local satellites
since their energy in redshifted in the satellite bandpass. Approximating
the energy spectrum as a power\,-\,law, i.e. $N(E) \propto E^{-\beta_{hr}}$ for 
large $E$, according to this scheme, high $z$ GRBs with more shallow profiles will 
likely be detected preferentially over the steeper one. As a consequence,
a plot of $\beta_{hr}$ vs $z$ should detect no correlation even if this is 
present. Schaefer (2007) reports the values of $\beta_{hr}$ for 28 of the 69 GRBs
in the calibrated HD, showing that there is indeed no variation with the redshift. 
For instance, roughly averaging over the 28 GRBs with measured $\beta_{hr}$, we find 
$\langle \beta_{hr} \rangle = -2.42$, while averaging over only the 17 GRBs 
with $z \le 1.4$ gives $\langle \beta_{hr} \rangle = -2.47$. Note, however, 
that both distributions are quite asymmetric with most of the GRBs having 
$\beta_{hr} \ge -2.7$ and only 3 with $\beta_{hr} \le -3$. It is also worth 
noting that, taken at face values, we find a clear correlation between $\log{L}$ 
and $\beta_{hr}$ with $\log{L}$ increasing with $\beta_{hr}$, although the 
large errors on $\beta_{hr}$ prevents from deeeming as statistically significant
such a trend. 

Quantifying the effect of a selection effect on $\beta_{hr}$ would need the
values of this parameter for all the GRBs in the sample which is not the case. 
However, since $\beta_{hr}$ and $\log{L}$ are positevely correlated, one can 
argue that selecting only GRBs with small $|\beta_{hr}|$ is the same as selecting 
those with the larger values of $\log{L}$ so that we can try investigating the
effects of a possible selection on $\log{L}$ as an alternative approach. Rather
than implementing a procedure based on the construction of mock samples (which
would also need the details of the detection strategies of the different 
satellites), we make an ideal experiment using the GRBs in our sample. As a first
step, we select only the GRBs with $z \le 1.4$ and divide them in three samples
referred to as CalAll, Cal51 and Cal52 containing all the GRBs with $\log{L}
\ge \lambda$ with $\lambda = 0, 51, 52$ respectively\footnote{Actually, the CalAll
and Cal51 samples are almost identical, but we have preferred to not furtherly divide
the sample because of too low statistics.}. Using these samples, we 
recalibrate the five 2D correlations using the local regression method described
in Sect. 4.2. We now use these calibrated relations to infer the GRBs distance
moduli whatever the value of $\log{L}$ is thus simulating the case of an 
unknown selection effect on $\log{L}$ being at work. Comparing the distance 
moduli thus obtained, we find\,:

\begin{displaymath}
\mu_{Cal51} = (1.00 {\pm} 0.03) \mu_{CalAll} + (-0.1 {\pm} 1.3) \ ,
\end{displaymath}

\begin{displaymath}
\mu_{Cal52} = (0.93 {\pm} 0.06) \mu_{CalAll} + (-2.6 {\pm} 2.7) \ ,
\end{displaymath}
while for the offset we find\,:

\begin{displaymath}
\langle \mu_{Cal51} - \mu_{CalAll} \rangle = 0.29 \pm 0.44 \ ,
\end{displaymath}

\begin{displaymath}
\langle \mu_{Cal52} - \mu_{CalAll} \rangle = 0.76 \pm 0.91 \ .
\end{displaymath}
It is worth stressing that the above linear fit has been done without
taking into account the errors on the variables thus giving all the points
the same weight. However, this qualitative analysis shows that, if a selection 
on $\log{L}$ should be at work, neglecting its effect does not bias the derived
distance moduli in a statistically meaningful way. On the other hand, there is 
actually an increase in $\Delta \mu$ as $\lambda$ increases thus motivating
further and more quantitative investigation on this issue through a careful set of simulations. 

\section{GRBs HDs as a dark energy probe}

The search for standard candles has been always motivated
by the need of building up reliable Hubble diagrams in order
to probe the evolution of the universe. As Eqs.(\ref{eq: defmu}) 
and (\ref{eq: cpldl}) clearly show, the Hubble diagram allows to
estimate the luminosity distance as a function of redshift and 
put constraints on the integrated Hubble parameter $H(z)$ and hence
on the cosmological parameters of a given model.  

The importance of SNeIa for such a project is impossible to underrate 
as it is witnessed by the discovery of the cosmic speed up and the 
overwhelming flood of papers using this dataset to test the viability of
different cosmological models. However, their main limitation is intimately
related to their nature preventing us from detecting SNeIa to redshift 
higher than $z \sim 2$. Moreover, even with future satellite missions, the
most of the sample will be made out of objects with $z \le 1$. As such, while 
they are excellent probes over the range $(0, 1)$ following the transition from
the dark energy dominated present day universe to a decellerating expansion, SNeIa
are unable to test the universe during the matter dominated era at $z \ge 2$. 
On the contrary, GRBs are ideal tools to probe the high redshift universe. That 
GRBs and SNeIa are complementary may be easily understood by considering, for instance,
the redshift distribution of the nowadays sample. Indeed, while SNeIa mainly covers
the range $(0, 1)$, GRBs are observed up to $z \sim 6.6$ with most of them 
lying at $z \ge 1$. As such, one may anticipate that they are less sensitive
to the details of the dark energy EoS since they mainly probe the high $z$ regime
deep in the matter dominated era. It is this peculiar feature, opposite to what
happens for SNeIa, that makes them a combined SNeIa\,+\,GRBs Hubble diagram
an ideal tool to constrain the universe expansion. 

\subsection{Present day data}

In order to show the power of a combined SNeIa\,+\,GRBs HD as a cosmological
probe, we present here a first analysis fitting this dataset to two 
popular models, namely the $\Lambda$CDM and CPL EoS. For both cases, the scaled 
Hubble parameter $E(z) = H(z)/H_0$ is given by Eq.(\ref{eq: ecpl} with 
the $\Lambda$CDM being obtained by setting $(w_0, w_a) = (-1, 0)$. In order to
constrain the model parameters, we maximize the following likelihood function\,:

\begin{equation}
{\cal{L}}({\bf p}) \propto \exp{(-\chi^2/2)} \ ,
\label{eq: deflikecosmo}
\end{equation}
with

\begin{eqnarray}
\chi^2({\bf p}) & = & \sum_{i = 1}^{{\cal{N}}_{SNeIa}}
{\left [ \frac{\mu_{obs}(z_i) - \mu_{th}(z_i, {\bf p})}
{\sigma_i} \right ]^2} \nonumber \\
~ & + &  \sum_{i = 1}^{{\cal{N}}_{GRB}}
{\left [ \frac{\mu_{obs}(z_i) - \mu_{th}(z_i, {\bf p})}
{\sqrt{\sigma_i^2 + \sigma_{int}^2}} \right ]^2} \nonumber \\
~ & + & \left ( \frac{h - h_{HST}}{\sigma_{HST}} \right )^2 
+ \left ( \frac{\omega_M - \omega_M^{CMBR}}{\sigma_M} \right )^2 \ .
\label{eq: defchicosmo}
\end{eqnarray}
In Eq.(\ref{eq: defchicosmo}), $\mu_{obs}$ and $\mu_{th}$ are the 
observed and theoretically predicted values of the distance modulus given
the set of model parameters ${\bf p}$, while the sum is over the ${\cal{N}}_{SNeIa}
= 307$ SNeIa in the Union dataset and the ${\cal{N}}_{GRB} = 69$ GRBs in
the calibrated HD. Note that, for GRBs, we have added to the denominator the 
constant error parameter $\sigma_{int}$ to take into account the intrinsic 
scatter\footnote{Note that a similar term is also present for SNeIa, but it is 
estimated to be $\sigma_{int} = 0.15$ and yet included into the error $\sigma_i$
provided in the dataset.} around the Hubble diagram due to the scatter in 
the 2D correlations. The last two terms in Eq.(\ref{eq: defchicosmo}) are 
Gaussian priors on $h$ and $\omega_M = \Omega_M h^2$ and are included in
order to help breaking the degeneracy with the other model parameters. To this 
aim, we resort to the result of the HST Key project \cite{Freedman} to set
$(h_{HST}, \sigma_{HST}) = (0.72, 0.08)$ and to the WMAP5 constraints 
\cite{WMAP5} giving $(\omega_M^{CMBR}, \sigma_M) = (0.138, 0.004)$. Note that
we have not added further data from other probes since we are here more 
interested in testing the ability of the combined HD to probe models rather
than setting strong constraints on the model themselves (which will be the 
subject of a forthcoming publication). 

In order to efficiently explore the parameter space, we run a Markov 
Chain Monte Carlo code generating chains with more than 100000 points 
which are enough to reach convergence. We then cut out the first $30\%$ of 
the chains to avoid the burn in phase and thineed the chains taking one
point each ten points in order to reduce fake correlations. For the 
$\Lambda$CDM model, we find for the best fit parameters\,:

\begin{displaymath}
(\Omega_M, h, \sigma_{int}) = (0.283, 0.701, 0.44)
\end{displaymath}
giving

\begin{displaymath}
\chi^2_{SNeIa}/d.o.f. = 1.03 \ \ , \ \ 
\chi^2_{GRB}/d.o.f. = 1.11 \ \ , \ \ 
\omega_M = 0.139 
\end{displaymath}
with $d.o.f. = {\cal{N}}_{data} - {\cal{N}}_{p}$ and ${\cal{N}}_p = 3$
for the $\Lambda$CDM case. The reduced $\chi^2$ close to $1$ for both
the SNeIa and GRB datasets and the very good agreement with the WMAP5 
$\omega_M$ value clearly shows that the concordance $\Lambda$CDM model 
still stands out as an excellent description of the universe expansion. 
After marginalizing over the other quantities, the constraints on the 
model parameters are found to be\,:

\begin{displaymath}
\Omega_M = 0.287_{-0.013 \ -0.029}^{+0.013 \ +0.027} \ , 
\end{displaymath}

\begin{displaymath}
h = 0.700_{-0.006 \ -0.012}^{+0.006 \ +0.012} \ ,
\end{displaymath}

\begin{displaymath}
\sigma_{int} = 0.15_{-0.11 \ -0.13}^{+0.21 \ +0.40} \ ,
\end{displaymath}
where, following the Bayesian prescription, we give the median  
as central value and the reported errors refer to the $68$ and $95\%$ CL. 
It is worth noting that the constraints on $\sigma_{int}$ shows that
this quantity is weakly constrained with the median values being very 
different from the best fit one. This is, however, not surprising. On the 
one hand, it is well known that, because of degeneracies and 
projection effects, the maximum of ${\cal{L}}({\bf p})$ is not given 
by the median values of the parameters. This could be the case only if 
the covariance matrix of the parameters is diagonal which is quite unusual. 
On the other hand, we have not included in the likelihood definition any prior
on $\sigma_{int}$ which could be inferred considering the intrinsic scatter 
of the 2D correlations used to build the GRB HD. Moreover, the asymmetry
in the $\sigma_{int}$ CL is a result of the obvious constraint that this quantity
must be positive because of its same defintion. 

As a further example, we now consider the CPL case leaving the 
parameters $(w_0, w_a)$ free in the fit thus giving ${\cal{N}}_p 
= 5$. For the best fit, we find\,:

\begin{displaymath}
(\Omega_M, w_0, w_a, h. \sigma_{int}) =
(0.273, -1.28, 2.02, 0.707, 0.45)
\end{displaymath}
giving 

\begin{displaymath}
\chi^2_{SNeIa}/d.o.f. = 1.03 \ \ , \ \ 
\chi^2_{GRB}/d.o.f. = 1.06 \ \ , \ \ 
\omega_M = 0.136 
\end{displaymath}
so that there is still a perfect agreement with the data. Note 
that, formally, the CPL model fits the GRB data better than 
the $\Lambda$CDM one, but the increase by two of the number of 
parameters is actually not compensated by a corresponding reduction
of the reduced $\chi^2_{GRB}$ so that a pure cosmological constant
is still preferred. One could be surprised that the best fit parameters
seem to favour a model which is very different from the $\Lambda$CDM one
with the present day value of the EoS deep in the phantom regime $w_0 < -1$
and a strongly varying EoS. However, this is just a consequence of the 
parameter degeneracies quoted above\footnote{Note also that we have
not set here the usual prior $w_0 + w_a = 0$ which is typically introduced 
to assure that the dark energy EoS fades away in the early universe.}. Indeed, 
the Bayesian constraints turn out to be\,:

\begin{displaymath}
\Omega_M = 0.283_{-0.022 \ -0.039}^{+0.021 \ +0.036} \ , 
\end{displaymath}

\begin{displaymath}
w_0 = -1.06_{-0.21 \ -0.42}^{+0.19 \ +0.34} \ , 
\end{displaymath}

\begin{displaymath}
w_a = 0.8_{-1.1 \ -1.9}^{+0.8 \ +1.8} \ , 
\end{displaymath}

\begin{displaymath}
h = 0.699_{-0.013 \ -0.022}^{+0.016 \ +0.029} \ ,
\end{displaymath}

\begin{displaymath}
\sigma_{int} = 0.08_{-0.05 \ -0.07}^{+0.15 \ +0.42} \ ,
\end{displaymath}
showing that the $\Lambda$CDM model with $(w_0, w_a) = (-1, 0)$ is well 
within the $68\%$ CL. It is worth noting that the constraints on both $w_0$
and $w_a$ are quite weak, as those on the intrinsic scatter $\sigma_{int}$. 
This same result is obtained if one only uses the SNeIa HD so that one could 
wonder whether adding the GRB data have improved the situation. Actually, this is
not becuase, as we have yet quoted above, the GRBs HD only probes the matter 
dominated era when the dark energy term is almost disappeared so that the 
precise values of $(w_0, w_a)$ are meaningless. As we will see later, GRBs improve,
on the other hand, the constraints on $\Omega_M$ even if this is not apparent for 
the fits we are considering here because of the addition of the priors on $h$ and $\omega_M$.  
As a final remark, we note that, even with these weak constraints, we can
exclude with high confidence historically discussed cosmological models
with, e.g., $(\Omega_M, \Omega_{\Lambda}) = (1, 0)$ or $(\Omega_M, \Omega_{\Lambda}) 
= (0.3, 0)$, and have a further strong evidence for an accelerating universe.

\subsection{Future SNeIa\,+\,GRB data}

The analysis of the present day data have not shown the full complemetarity
between SNeIa and GRBs, essentially because we have added the priors on the 
two parameters $h$ and $\omega_M$. To better investigate this issue, we compute 
the Fisher information matrix (see, e.g., Tegmark et al. 1997 and references therein) 
defined as\,:

\begin{equation}
F_{ij} = \left \langle \frac{\partial^2 L}{\partial p_i \partial p_j} \right \rangle
\label{eq: deffij}
\end{equation}
with $L = -2 \ln{{\cal{L}}({\bf p})}$, and we redefine the likelihood as in 
Eq.(\ref{eq: deflikecosmo}), but excluding the two priors on $(h, \omega_M)$. 
Note that $\langle \ldots \rangle$ denotes the expectation value, but, actually, 
this is computed by evaluating the Fisher matrix elements for fiducial values of 
${\bf p}$. The inverse of the Fisher matrix gives the covariance matrix ${\bf C}$ 
so that $\sigma_i = \sqrt{C_{ii}}$ is an estimate of the error on the parameter $i$. 
We remember that, because of the Cramer\,-\,Rao theorem, this is an upper limit 
on the error on the parameter $p_i$ that can be obtained by an experiment with 
the given characteristics. 

A key ingredient in the computation of ${\bf F}$ is represented by the 
details of the SNeIa and GRB surveys giving the redshift distribution of the samples
and the errors on each measurement. For SNeIa, following Kim et al. (2004), we 
adopt\footnote{Note that, in Kim et al. (2004), the authors assume the data are
separated in redshift bins so that the error becomes $\sigma^2 =
\sigma_{sys}^2/{\cal{N}}_{bin} + {\cal{N}}_{bin} (z/z_{max})^2 \sigma_m^2$
with ${\cal{N}}_{bin}$ the number of SNeIa in a bin. However, we prefer to
not bin the data so that ${\cal{N}}_{bin} = 1$.}\,:

\begin{displaymath}
\sigma(z) = \sqrt{\sigma_{sys}^2 + \left ( \frac{z}{z_{max}} \right )^2 \sigma_m^2}
\end{displaymath}
with $z_{max}$ the maximum redshift of the survey, $\sigma_{sys}$ an
irreducible scatter in the SNeIa distance modulus and $\sigma_m$ to be
assigned depending on the photometric accuracy. For GRBs, we assume that the
redshift distribution is provided is the same as the star formation history in 
Porciani \& Madau (2001) although it is worth noting that this just a first order 
approximation. Indeed, one should convolve this theoretical distribution with
the detection efficiency of a satellite, like {\it Swift} or {\it Fermi}, which will
also claim for a model of GRB explosion to compute the expected fluence. Since all
these quantities are up to know largely uncertain, we prefer to not detail them warning
the reader that our results are somewhat optimistic. Concerning the error
distribution, we assume that the main source of error is the intrinsic scatter (as it is
indeed the case with the present day data) so that we give to each GRBs distance 
modulus an uncertainty randomly generated from a Gaussian centred on $\sigma_{GRB}$ and 
with standard deviation $\sigma_{std}$. To further concentrate our attention
on the constraints on the cosmological parameters, we moreover assume that
the intrinsic scatter $\sigma_{int}$ in Eq.(\ref{eq: defchicosmo}) has been
estimated somewhat and included in the error on $\mu$ so that this is no more 
a quantity to be constrained. Finally, in order to run the Fisher matrix calculation, 
we have to choose a fiducial model which we do by setting $(\Omega_M, w_0, w_a, h) = 
(0.277, -1, 0, 0.72)$ in agreement with the WMAP5 results.

As a first test, we consider the present day data thus using the observed 
redshift distribution as the Union one for SNeIa and that of the calibrated HD
for GRBs. We further set $({\cal{N}}_{SNeIa}, \sigma_m) = (307, 0.33)$ and 
$(\sigma_{GRB}, \sigma_{STD}) = (0.58, 0.22)$. With these values, we get\,:

\begin{displaymath}
\sigma(\Omega_M) = 0.58 \ \ , \ \ \sigma(w_0) = 0.17 \ , 
\ \ \sigma(w_a) = 1.2 \ \ ,
\end{displaymath}
using SNeIa data only, while adding GRBs gives\,:

\begin{displaymath}
\sigma(\Omega_M) = 0.08 \ \ , \ \sigma(w_0) = 0.08 \ \ , 
\ \ \sigma(w_a) = 0.4 \ \ .
\end{displaymath}
Such a test immediately shows the complementarity of SNeIa and GRBs with
these latter data halving the error on $\Omega_M$ thanks to their ability
of probing the matter dominated era. Note also the reduction of the errors 
on $(w_0, w_a)$ which seems to contradict our claim that GRBs are not 
sensitive to the dark energy EoS. Actually, the reduction is due
mainly to GRBs increasing the statistics rather than probing the EoS as 
is better understood considering future data. 

To this end, we now simulate a SNAP\,-\,like SNeIa survey setting $({\cal{N}}_{SNeIa}, 
\sigma_m) = (2000, 0.02)$ and using the Porciani \& Madau (2001) distribution
to generate 200 GRBs leaving the error parameters $(\sigma_{GRB}, \sigma_{std})$ 
unchanged. Using only SNAP SNeIa, we get\,:

\begin{displaymath}
\sigma(\Omega_M) = 0.077 \ \ , \ \ \sigma(w_0) = 0.039 \ , 
\ \ \sigma(w_a) = 0.22 \ \ ,
\end{displaymath}
while adding GRBs gives\,:

\begin{displaymath}
\sigma(\Omega_M) = 0.019 \ \ , \ \sigma(w_0) = 0.036 \ \ , 
\ \ \sigma(w_a) = 0.20 \ \ .
\end{displaymath}
The role of the complementarity of SNeIa and GRBs is now better 
explained. Indeed, while $\sigma(\Omega_M)$ is reduced by a factor 
$\sim 4$, both $\sigma(w_0)$ and $\sigma(w_a)$ are almost unaltered
by the addition of GRBs in agreement with the claim that their Hubble 
diagram is unable to put constraints on $(w_0, w_a)$.

As a final remark, it is worth noting that the detection of 200 GRBs is 
likely well within the capabilities of both the {\it Swift} and {\it Fermi}
satellites so that one can argue that, by the time the SNAP survey will be 
completed, one indeed has the opportunity to build up a combined SNeIa\,+\,GRB
HD as the one assumed in our Fisher matrix forecast. However, it is worth
stressing that the precision attainable with GRBs may critically depend on their
assumed redshift distribution and on the intrinsic scatter of the 2D correlations
used to build up the calibrated HD. Both these key ingredients have been 
modelled very roughly in our analysis so that a more careful investigation should
be needed. Although such a work can downgrade our optimistic forecast, we argue 
that similar results can still be obtained by increasing the number of GRBs still
remaining within the realm of the {\it Fermi} satellite detection capabilities.

\section{Conclusions}

As soon as the need to probe the Hubble diagram to a redshift higher than
the one attainable with SNeIa became clear, most attention has been devoted
to search for a method to make GRBs standardizeable candles. To this aim,
different 2D correlations between a luminosity or energy distance dependent
quantity and observationally accessible GRBs properties have been proposed
in order to build up a GRBs Hubble diagram to be used for constraining
cosmological parameters. In an attempt to make a step forward along this
road, we have here reconsidered different aspects of the problem trying
with the final aim of constructing a more reliable GRBs Hubble diagram.

First, we have added the recently found $L_X$\,-\,$T_a$ correlation to the
other five 2D empirical laws used up to now. Although being different from
the other ones since it is based on afterglow rather than peak quantities,
we have shown that the estimated distance modulus for each of the GRBs
common to the Schaefer (2007) and Willingale et al. (2007) samples is in
agreement with the one obtained using the standard set of 2D correlations.
As such, one can be confident that no systematic bias is introduced by
resorting to this new empirical law. On the contrary, its use makes it
possible to both reduce the errors on $\mu(z)$ by a significative $\sim
14\%$ and increase the sample from 69 to 83 GRBs. With the {\it Swift}
satellite still operating and collecting data on the X\,-\,ray afterglow,
the prospects to both improve the estimate of the calibration parameters
and adding still more GRBs to the sample are quite good. Needless to say,
this is also possible for the peak based 2D correlations. It is worth
noting that, in order to reduce the uncertainty on $\mu(z)$, following
Schaefer (2007), we have made a weighted average of the estimates coming
from the use of the six 2D correlations. While this makes it possible to,
in a sense, average out any systematics intrinsic to each particular
correlations, such a procedure is not statistically well motivated.
Actually, comparing the estimated $\mu(z)$ with each other and with the
weighted average makes us confident that this method have not biased anyway
the final results. However, increasing the sample will allow us to better
cross check the results thus leading further confidence in this approach.

As a preliminary step in the construction of the GRBs HD, we have updated
the calibration of the six 2D correlations we have used. To this aim, we
have first to assume a background cosmological model to compute the
distance dependent quantities. The recent WMAP5 data analysis has motivated
our choice of a fiducial flat $\Lambda$CDM model with $(\Omega_M, h) =
(0.291, 0.697)$, but we have also explored the effect of varying both
$(\Omega_M, h)$ and the $(w_0, w_a)$ parameters of the phenomenological CPL
dark energy EoS. These data are then used as input to a Bayesian motivated
fitting procedure which makes it possible to estimate both the slope $a$
and the intrinsic scatter $\sigma_{int}$ of the 2D correlations correctly
taking into account the errors on both involved variables. To our
knowledge, this is the first time that the 2D GRBs empirical laws have been
calibrated using the same statistical tool for all of them. Moreover, our
method avoids to underestimate the intrinsic scatter which plays an
important role not only in the determination of the distance modulus
uncertainty, but also in testing theoretical models of GRBs explosion and
afterglow. As a somewhat surprising result, we find that the calibration
parameters $(a, b, \sigma_{int})$ very weakly depend on the assumed
cosmological model, i.e. on the value of $(\Omega_M, h, w_0, w_a)$. As a
consequence, the fiducial GRBs HD (i.e., the one relying on the
calibrations made in the fiducial $\Lambda$CDM model) is essentially
unaffected by varying the cosmology. Although this nice finding argues in
favour of the use of the fiducial HD as a tool to constrain different
cosmological models, we caution the reader that we have only explored a
single class of dark energy theories. Indeed, the CPL parametrization makes
it possible to mimic a quite large set of models, but may fail to mimic
scenarios with a significative quantity of dark energy at high redshift. It
is likely that the impact on $(a, b, \sigma_{int})$ is larger in this class
of theories, even if it is worth stressing that they are disfavoured by the
present dataset and the constraints from primordial nucleosynthesis.

In order to fully avoid the problem of assuming any background cosmology in
the calibration procedure, we have illustrated a novel method relying on
the estimate of the distance modulus of a given GRBs at redshift $z$
through a local regression analysis of the Union SNeIa sample. As we have
convincingly shown, local regression allows to recover the right value of
$\mu(z)$ whatever are $z$ and the parameters $(\Omega_M, h, w_0, w_a)$ of
the underlying CPL cosmology. As a consequence, we can calibrate the 2D
correlations using GRBs with $z \le 1.4$ (i.e., overlapping the redshift
range probed by SNeIa) without the need of assuming any fiducial
cosmological model. This procedure finally leads us to what we term the
calibrated GRBs HD containing 69 objects with model independent estimates
of the distance modulus. Reassuringly, for the GRBs in common, the
calibrated and fiducial HDs are in very good agreement thus suggesting that
both of them are not affected by any systematic bias induced by the
different calibration procedures.

One can wonder which is the GRBs HD better suited as a cosmological tool.
This is somewhat a matter of personal taste. Indeed, the two HDs agree in
all of their properties also providing the same percentage of likely
outliers with respect to a fiducial $\Lambda$CDM model. On one hand, the
fiducial HD contains a larger number of objects with smaller uncertainties,
but, although this is unlikely, one can still not exclude a bias
originating from the model dependent calibration. On the other hand, the
calibrated HD is free of any problem due to the unknown background
cosmology, but its constraining power is likely to be weaker because of the
lower number of objects and the greater uncertainties, both effects due to
not having used the $L_X$\,-\,$T_a$ correlation. Should future data make it
possible to calibrate also the $L_X$\,-\,$T_a$ law with the local
regression based method, the calibrated HD should become as precise as the
fiducial HD. As a concluding remark, we therefore argue in favour of this
latter approach as the best way to use GRBs as cosmological tools.

\section*{Acknowledgements}

We sincerely thank an anonymous referee for his/her comments which have 
helped to improve the paper. VFC is supported by Regione 
Piemonte and Universit\`a di Torino. Partial support from INFN project PD51 is 
acknowledged too. 

\appendix

\section{Hubble diagrams data}

\begin{table}
\caption{Fiducial and calibrated Hubble diagrams data for the 69 GRBs
common to both samples. Order of the columns as follows\,: 1. GRBs id; 2.
redshift; 3., 4. distance moduli (with $1 \sigma$ error) from the fiducial
and calibrated HDs.}
\begin{center}
\begin{tabular}{|c|c|c|c|}
\hline
Id & $z$ & $\mu_{fid}$ & $\mu_{cal}$ \\ \hline
\hline
970228 & 0.70 & 43.04 ${\pm}$ 0.68 & 42.72 ${\pm}$ 0.68 \\ 970508 & 0.84 & 43.79
${\pm}$ 0.44 & 43.76 ${\pm}$ 0.35 \\ 979828 & 0.96 & 43.38 ${\pm}$ 0.49 & 43.07 ${\pm}$
0.38 \\ 971214 & 3.42 & 47.77 ${\pm}$ 0.59 & 47.54 ${\pm}$ 0.59 \\ 980613 & 1.10 &
44.93 ${\pm}$ 1.29 & 44.75 ${\pm}$ 1.22 \\ 980703 & 0.97 & 44.16 ${\pm}$ 0.42 & 43.84
${\pm}$ 0.32 \\ 990123 & 1.61 & 44.36 ${\pm}$ 0.47 & 44.66 ${\pm}$ 0.37 \\ 990506 &
1.31 & 43.94 ${\pm}$ 0.56 & 43.76 ${\pm}$ 0.53 \\ 990510 & 1.62 & 45.31 ${\pm}$ 0.42 &
45.36 ${\pm}$ 0.31 \\ 990705 & 0.84 & 43.72 ${\pm}$ 0.49 & 43.40 ${\pm}$ 0.38 \\ 990712
& 0.43 & 41.96 ${\pm}$ 0.62 & 41.76 ${\pm}$ 0.44 \\ 991208 & 0.71 & 41.87 ${\pm}$ 0.66
& 41.65 ${\pm}$ 0.65 \\ 991216 & 1.02 & 42.95 ${\pm}$ 0.44 & 43.12 ${\pm}$ 0.35 \\
000131 & 4.50 & 47.21 ${\pm}$ 0.68 & 47.14 ${\pm}$ 0.68 \\ 000210 & 0.85 & 42.34
${\pm}$ 0.66 & 42.27 ${\pm}$ 0.65 \\ 000911 & 1.06 & 44.21 ${\pm}$ 0.68 & 44.27 ${\pm}$
0.67 \\ 000926 & 2.07 & 45.39 ${\pm}$ 0.69 & 45.09 ${\pm}$ 0.68 \\ 010222 & 1.48 &
43.98 ${\pm}$ 0.46 & 44.62 ${\pm}$ 0.29 \\ 010921 & 0.45 & 43.00 ${\pm}$ 0.58 & 42.53
${\pm}$ 0.54 \\ 011211 & 2.14 & 45.77 ${\pm}$ 0.62 & 45.53 ${\pm}$ 0.43 \\ 020124 &
3.20 & 46.88 ${\pm}$ 0.45 & 46.73 ${\pm}$ 0.37 \\ 020405 & 0.70 & 43.48 ${\pm}$ 0.52 &
43.47 ${\pm}$ 0.46 \\ 020813 & 1.25 & 44.07 ${\pm}$ 0.43 & 43.95 ${\pm}$ 0.33 \\ 020903
& 0.25 & 43.07 ${\pm}$ 1.26 & 42.23 ${\pm}$ 1.16 \\ 021004 & 2.32 & 46.81 ${\pm}$ 0.52
& 46.60 ${\pm}$ 0.48 \\ 021211 & 1.01 & 43.94 ${\pm}$ 0.61 & 43.49 ${\pm}$ 0.55 \\
030115 & 2.50 & 46.60 ${\pm}$ 0.59 & 46.25 ${\pm}$ 0.57 \\ 030226 & 1.98 & 46.66
${\pm}$ 0.48 & 46.50 ${\pm}$ 0.40 \\ 030323 & 3.37 & 47.79 ${\pm}$ 0.97 & 47.65 ${\pm}$
0.96 \\ 030328 & 1.52 & 44.80 ${\pm}$ 0.50 & 44.68 ${\pm}$ 0.37 \\ 030329 & 0.17 &
40.56 ${\pm}$ 0.42 & 39.73 ${\pm}$ 0.29 \\ 030429 & 2.66 & 46.80 ${\pm}$ 0.58 & 46.61
${\pm}$ 0.53 \\ 030528 & 0.78 & 45.12 ${\pm}$ 0.58 & 44.31 ${\pm}$ 0.54 \\ 040924 &
0.86 & 44.06 ${\pm}$ 0.61 & 43.61 ${\pm}$ 0.55 \\ 041006 & 0.71 & 44.03 ${\pm}$ 0.50 &
43.92 ${\pm}$ 0.42 \\ 050126 & 1.29 & 46.33 ${\pm}$ 0.55 & 45.74 ${\pm}$ 0.52 \\ 050318
& 1.44 & 46.09 ${\pm}$ 0.51 & 45.95 ${\pm}$ 0.44 \\ 050319 & 3.24 & 47.53 ${\pm}$ 0.66
& 47.73 ${\pm}$ 0.93 \\ 050401 & 2.90 & 46.15 ${\pm}$ 0.53 & 45.94 ${\pm}$ 0.55 \\
050406 & 2.44 & 48.44 ${\pm}$ 0.72 & 48.03 ${\pm}$ 0.70 \\ 050408 & 1.24 & 45.52
${\pm}$ 0.76 & 45.09 ${\pm}$ 0.72 \\ 050416 & 0.65 & 43.98 ${\pm}$ 0.85 & 43.32 ${\pm}$
0.81 \\ 050502 & 3.79 & 47.51 ${\pm}$ 0.63 & 47.24 ${\pm}$ 0.64 \\ 050505 & 4.27 &
48.47 ${\pm}$ 0.55 & 48.49 ${\pm}$ 0.59 \\ 050525 & 0.61 & 43.55 ${\pm}$ 0.44 & 43.28
${\pm}$ 0.37 \\ 050603 & 2.82 & 45.12 ${\pm}$ 0.55 & 44.66 ${\pm}$ 0.58 \\ 050802 &
1.71 & 45.50 ${\pm}$ 0.66 & 45.52 ${\pm}$ 0.98 \\ 050820 & 2.61 & 46.51 ${\pm}$ 0.63 &
46.27 ${\pm}$ 0.59 \\ 050824 & 0.83 & 44.65 ${\pm}$ 1.19 & 44.07 ${\pm}$ 1.19 \\ 050904
& 6.29 & 48.86 ${\pm}$ 0.52 & 49.27 ${\pm}$ 0.47 \\ 050908 & 3.35 & 47.38 ${\pm}$ 0.76
& 47.00 ${\pm}$ 0.76 \\ 050922 & 2.20 & 45.71 ${\pm}$ 0.55 & 45.57 ${\pm}$ 0.52 \\
051022 & 0.80 & 43.61 ${\pm}$ 0.46 & 43.77 ${\pm}$ 0.28 \\ 051109 & 2.35 & 45.87
${\pm}$ 0.83 & 45.84 ${\pm}$ 0.80 \\ 051111 & 1.55 & 45.14 ${\pm}$ 0.63 & 44.54 ${\pm}$
0.60 \\
\hline
\end{tabular}
\end{center}
\end{table}

\begin{table}
\contcaption{}
\begin{center}
\begin{tabular}{|c|c|c|c|}
\hline
Id & $z$ & $\mu_{fid}$ & $\mu_{cal}$ \\ \hline
\hline
060108 & 2.03 & 47.93 ${\pm}$ 0.72 & 48.85 ${\pm}$ 1.07 \\ 060115 & 3.53 & 48.21
${\pm}$ 0.80 & 47.78 ${\pm}$ 0.79 \\ 060116 & 6.60 & 48.67 ${\pm}$ 0.77 & 48.33 ${\pm}$
0.92 \\ 060124 & 2.30 & 46.35 ${\pm}$ 0.40 & 46.78 ${\pm}$ 0.38 \\ 060206 & 4.05 &
46.55 ${\pm}$ 0.51 & 46.37 ${\pm}$ 0.60 \\ 060210 & 3.91 & 48.30 ${\pm}$ 0.45 & 48.59
${\pm}$ 0.47 \\ 060223 & 4.41 & 47.94 ${\pm}$ 0.57 & 47.64 ${\pm}$ 0.54 \\ 060418 &
1.49 & 45.09 ${\pm}$ 0.47 & 45.00 ${\pm}$ 0.51 \\ 060502 & 1.51 & 45.37 ${\pm}$ 0.65 &
44.90 ${\pm}$ 0.62 \\ 060510 & 4.90 & 48.66 ${\pm}$ 0.94 & 48.60 ${\pm}$ 0.93 \\ 060526
& 3.21 & 47.56 ${\pm}$ 0.43 & 47.17 ${\pm}$ 0.41 \\ 060604 & 2.68 & 47.18 ${\pm}$ 0.55
& 46.23 ${\pm}$ 0.57 \\ 060605 & 3.80 & 47.66 ${\pm}$ 0.57 & 47.04 ${\pm}$ 0.68 \\
060607 & 3.08 & 46.68 ${\pm}$ 0.58 & 46.24 ${\pm}$ 0.55 \\
\hline
\end{tabular}
\end{center}
\end{table}

\begin{table}
\caption{Hubble diagram data for the 14 GRBs present in the fiducial one only. Order
of the columns as in Table A1.}
\begin{center}
\begin{tabular}{|c|c|c|}
\hline
Id & $z$ & $\mu_{fid}$ \\ \hline
\hline
050315 & 1.95 & 46.11 ${\pm}$ 0.98 \\ 050416A & 0.65 & 44.73 ${\pm}$ 1.14 \\
050820A & 2.61 & 45.43 ${\pm}$ 0.97 \\ 050922C & 2.20 & 44.74 ${\pm}$ 0.98 \\
051016B & 0.94 & 42.81 ${\pm}$ 1.65 \\ 051109A & 2.35 & 45.49 ${\pm}$ 1.17 \\
060223A & 4.41 & 50.09 ${\pm}$ 1.25 \\ 060502A & 1.51 & 46.06 ${\pm}$ 1.36 \\
060522 & 5.11 & 49.65 ${\pm}$ 1.10 \\ 060607A & 3.08 & 49.43 ${\pm}$ 1.73 \\ 060614
& 0.12 & 42.26 ${\pm}$ 0.88 \\ 060707 & 3.43 & 48.12 ${\pm}$ 1.41 \\ 060714 & 2.71
& 45.41 ${\pm}$ 1.10 \\ 060729 & 0.54 & 42.55 ${\pm}$ 0.89 \\
\hline
\end{tabular}
\end{center}
\end{table}

As explained in the main text, the fiducial and calibrated HDs contain a
different number of GRBs because it is not possible to use the local
regression calibration for the $L_X$\,-\,$T_a$ relation. Therefore, we
report in Table A1 the redshift and distance moduli of the 69 GRBs common
to both samples, while Table A2 gives the same data for the 14 GRBs present
only in the fiducial HD. Further material (such as the values of $\mu$ from
the single 2D correlations) is available on request.

\section{Comments on outliers}

As we have quoted in Sect. 4.3, there are 13 (10) GRBs deviating more than
$2\sigma$ from the fiducial (calibrated) HD which are therefore likely
outliers. Actually, dealing with {\it likely outliers} is a complicated
task since one has to find a serious motivation before rejecting a given
object from a sample. To this end, one should look both at its eventually
peculiar features and the instrumental and observing setup which has been
used to gather the corresponding data. Since we have no access to these
informations for all GRBs, we have preferred to not reject them.

Nonetheless, as an example, we discuss here the case of GRB\,060614 which
is the only GRB deviating more than $3\sigma$ from both the fiducial HD and
the $L_X$\,-\,$T_a$ correlation. Indeed, this object is somewhat very
peculiar being the first long GRB observed at low redshift ($z = 0.125)$
and clearly not associated to a bright Ib/c SN \cite{DellaValle06}.
Actually, its short duration ($T_{90} = 100 \ {\rm s}$) makes it hardly
classifiable as a long GRB, while its morphology is quite similar to the
one of GRB\,0507024 which is indeed classified as a short GRB
\cite{Piro05,Zhang07}. It is worth noting that GRB\,060614 has been
interpreted both in the framework of the canonical fireshell model
\cite{Ruffini01,Caito09} and by \cite{NB06} as a new class of intermediate
sources. All these peculiarities indeed argue in favour of the rejection of
this GRB from the fiducial HD as a physical outlier.

Another interesting case is represented by GRB\,060505 which is not
associated by any SN thus suggesting it is representative of a class of
objects with different progenitors. However, among the likely outliers, we
also find GRB\,030329 wich is one of spectroscopically confirmed GRBs
associated with a SN. However, its X\,-\,ray afterglow was observed with
only two pointings of the Rossi\,-\,XTE instrument obtained respectivly 5
hours and 1.24 days after the burst \cite{MS03,Marsh03}. Actually, an
incomplete coverage of the X\,-\,ray lightcurve typically takes place for
many GRBs observed with the old missions such as Beppo SAX, Rossi, Hete2.
One can argue that the parameters obtained in these cases are somewhat
uncorrectly estimated thus altering the position of the corresponding GRB
in the fiducial HD. This could be the case for 10 out of 13 likely ouliers
GRBs (namely\,: 990123, 991208, 991216, 990506, 000210, 010222, 020813,
020903, 030329, 030528) thus suggesting that they could be rejected because
of observationally (rather than physically) motivated problems.

As it should be clear from these comments, we have up to now only some
phenomenological and qualitative hints to address the problem of outliers.
We therefore deserve a detailed analysis also taking into account more
quantitative features to a forthcoming publication.

\end{document}